# Laser-Assisted Multilevel Non-Volatile Memory Device Based on 2D van-der-Waals Few-layer-ReS$_2$/h-BN/Graphene Heterostructures

*Bablu Mukherjee*\*, *Amir Zulkefli, Kenji Watanabe, Takashi Taniguchi, Yutaka Wakayama, Shu Nakaharai*\*

Dr. B. Mukherjee, A. Zulkefli, Prof. Y. Wakayama, Dr. S. Nakaharai

International Center for Materials Nanoarchitectonics (MANA),

\* Corresponding authors. Email: MUKHERJEE.Bablu@nims.go.jp (B. Mukherjee);

NAKAHARAI.Shu@nims.go.jp (S. Nakaharai)

A. Zulkefli, Prof. Y. Wakayama

Department of Chemistry and Biochemistry, Faculty of Engineering, Kyushu University,

Dr. K. Watanabe, Dr. T. Taniguchi

Research Center for Functional Materials,

National Institute for Materials Science (NIMS), 1-1 Namiki, Tsukuba, Ibaraki 305-0044,

Japan.

Few-layer ReS$_2$ field-effect transistors (FET) with a local floating gate (FG) of monolayer graphene separated by a thin h-BN tunnel layer for application to a non-volatile memory (NVM) device is designed and investigated. FG-NVM devices based on two-dimensional (2D) van-der-Waals (vdW) heterostructures have recently been studied as important components to realize digital electronics and multifunctional memory applications. Direct bandgap multilayer $ReS_2$ satisfies various requirements as a channel material for electronic devices as well as being a strong light-absorbing layer, which makes it possible to realize light-assisted optoelectronic applications. The non-volatile memory operation with a high ON/OFF current ratio, a large memory window, good endurance (> 1000 cycles) and stable retention (> $10^4$ s) have been observed. We demonstrate successive program and erase states using 10 millisecond gate pulses of +10 V and -10 V, respectively. Laser pulses along with electrostatic gate pulses provide multi-bit level memory access via opto-electrostatic coupling. The devices exhibit the dual functionality of a conventional electronic memory and can store laser-pulse excited signal information for future all-optical logic and quantum information processing.

## 1. Introduction

The atomically flat two-dimensional (2D) surfaces of layered materials make it possible to realize tuneable properties via strong electrostatic coupling. A van-der-Waals (vdW) 2D heterostructure combining a semiconductor, an insulator and a conductor has various important applications in digital electronics and memory operations. [1-3] For instance, insulating atomically flat hexagonal boron nitride (h-BN) can function as a tunnelling layer between a conductor and a semiconducting channel thus allowing perfectly planar charge injection across the atomically flat interface. [4,5] Recently non-volatile memories (NVM) based on 2D heterostructures and their tuneable and multibit operation under laser light have been reported. [6-15] Laser-assisted memory operation provides a new degree of freedom for multifunctional optoelectronic devices with the extra functionality of optically communicated multilevel access. [1,16-18] Laser light can travel through free space without losing power and this allows us to operate optoelectronic devices from a long distance away at low power and with little need for maintenance. Therefore, there is a need for a low operating power and optical control through fibre cable for the multi-level operation of NVM devices. In addition to the data storage capability, there is a strong need for an NVM optoelectronic device that can distinguish light wavelengths for colour sensing in digital imaging. [19]

FG-NVM devices based on 2D vdW heterostructures have recently been studied as important components with which to develop digital electronics and optoelectronics. Compared with conventional Si 3D semiconductors, all-2D layered materials and their heterostructures are superior for NVM applications due to their atomically flat surfaces without dangling bonds, high carrier mobility, vdW interaction in a vertically stacked geometry and the potential for increasing the integration density. Direct bandgap multilayer ReS$_2$ satisfies various requirements as a channel material for memory device applications since it exhibits excellent electrical properties with high carrier mobility compared with other transition metal dichalcogenides (TMDCs) such as MoS$_2$. Few-layer ReS$_2$ material is a good channel material for field-effect transistors (FET). Moreover, it is a good light absorbing material thanks to its thickness-independent direct band gap ( ~ 1.4 – 1.5 eV), which is useful for optics and optoelectronics applications. [20-23] A thin layer of h-BN can be used to further enhance field-effect mobility due to reduce charge trapping/scattering sites and atomically flat electrostatic modulation. BN is used as a thin insulating tunnel layer, and highly conductive monolayer graphene (MLG) can be used as a floating gate (FG) that can store charge carriers for device applications. Despite the outstanding electrical, optical and optoelectrical properties of this device, which consists of a vertical heterostack with a semiconductor/insulator/conductor structure formed of ReS$_2$/h-BN/FG-graphene, there is still a need to realize a multibit photo-signalling memory structure that operates at low power, with a high ON/OFF current ratio and a large memory window for all-optical logic processing devices.

The main purpose of this work is to evaluate the device performance of multilayer ReS$_2$ FETs with a local FG and to investigate the effects of a graphene floating gate when separated with a thin h-BN tunnel layer, for application to a multilevel NVM device via opto-electrostatic coupling. The proposed photoelectronic memory of multilayer-ReS$_2$/h-BN/FG-graphene/SiO$_2$ devices can be operated in the same way as a conventional electronic memory at a low programming gate voltage. Moreover, the device can store laser pulse information as an electronic readout. In addition to offering dual functionality, a memory device with a large hysteresis window could open the way to achieving multibit data storage via both electrostatic gate pulses and incident laser pulses. In this work, we demonstrate the first NVM operation via electrostatic coupling. The output characteristics ($I_d$-$V_d$) under a constant gate bias and different gate bias pulse widths indicate a large hysteresis, which is related to the charge accumulation in FG-graphene as the result of tunnelling through thin h-BN. We investigated low power operation and a large memory window for multi-bit storage applications. The drain current-time characteristics ($I_d$-$t$) at positive (+10 V) and negative (-10 V) gate pulse biases, respectively, make it possible to access two different states, namely 'Program' and 'Erase', during NVM operation with a fast reading/writing speed (of the order of milliseconds). We demonstrate successive store/erase operations corresponding to logic `0` and `1`, respectively, via a cycle of positive/negative electrostatic gate pulses. In the second part of this work, we characterize the device performance under light illumination with laser wavelength of 532 nm, whose energy is higher than ReS$_2$ direct bandgap energy. Multi-level memory operation was demonstrated using opto-electrostatic coupling, in which the device communicated by using laser pulses along with the gate pulse dependence of the 'ON' and 'OFF' states of the memory device. Also, incident laser coupling at different powers allowed us to store multi-bit memory states. In addition to laser-power-dependent multi-level memory operation, we demonstrate memory operation by irradiating fixed power laser pulses and programmed gate pulses, where the device readout electronic memory state depends on the incident laser pulse frequency and electrostatic gate pulse frequency. This study provides an overview of the use of all-2D atomically-flat layered materials to realize various electronic components including sensors, detectors and memory devices for the Internet-of-Things (IoT) era. The photonic signal storage that we propose in this work can potentially enable all-optical logic processing and quantum information processing.

**2. Results and Discussions**

Our field-effect transistor (FET) structure for NVM application is a vdW heterostructure consisting of a channel layer of multilayer $ReS_2$ supported by high-quality thin h-BN as a carrier tunnelling layer, where we placed monolayer graphene beneath the h-BN as a floating gate to store charge carriers. Optical images of the different components used to fabricate the heterostructure and the electrical contacts used to form the device are shown in **Figure 1 a, b**, respectively. The flake transfer method that we employed to form the heterostructure of the NVM device is described in the Methods section and the Supporting Information (**Figure S1**). The transfer method involves the deterministic transfer of large-area high-quality 2D flakes onto an arbitrary substrate. Three-dimensional (3D) and cross-sectional views of a typical device architecture are shown in **Figure 1c, d**, respectively. We applied a back-gate voltage (control gate bias $V_g$) to the degenerately doped silicon substrate ($p^{++}Si$) to tune the memory characteristics via electrostatic coupling and the systematic shifting of the Fermi level of the $ReS_2$ channel along with band alignment across the $ReS_2$/h-BN/FG-graphene heterostack. Before testing the electrical characteristics of the heterostructure FETs, we tested the electrical performance of graphene, $ReS_2$ and $ReS_2$/h-BN FET devices. **Figure S2 (a, b)** shows the *p*-doped electrical characteristics of a monolayer graphene FET that result from the absorption of oxygen or water in air. [24] Here, a gate voltage of +20 V is needed to reach the charge neutrality point. Graphene was immediately transferred to the top of the treated $SiO_2$/Si substrate, and the result was that there was almost no hysteresis in the sweep $I_d$-$V_g$ characteristics due to the lack of water molecules at the graphene/SiO$_2$ interface. **Figure S2 (c, d)** shows the electrical performance of ReS$_2$ FETs with and without a bottom h-BN layer. The field mobility of the device was enhanced because there were fewer charge trapping/scattering sites and an atomically flat electrostatic modulation was realized by using a thin layer of h-BN as the tunnel layer of the device.

We confirmed the crystalline quality of the transferred ReS$_2$ layer by Raman spectroscopy (**Figure 2a**), which indicated the highly crystalline and high chemical purity of the ReS$_2$ flake material. The ReS$_2$ layer was a few atomic layers (~5 nm) thick, which was further verified with an AFM scan and a line profile monitoring the thickness (**Figure 2d**). We observed typical E$_{2g}$ (in-plane vibration mode) and A$_{1g}$ (out-of-plane vibration mode) peaks at 162 and 212 cm$^{-1}$, respectively, along with other labelled peaks. The Raman spectrum of graphene (**Figure 2b**) exhibits a highly crystalline monolayer thickness signature as it has a 2D/G peak integral intensity ratio greater than 2 and a narrow (and symmetric) 2D peak width (FWHM) of 22.1 ± 0.5 cm$^{-1}$ at 2680.4 cm$^{-1}$. Other Raman peak signatures were present, namely those of an E$_{2g}$ in-plane phonon caused by the stretching C-C vibration mode and of a D peak at 1350 cm$^{-1}$, which is related to defects. The h-BN signature was observed in the E$_{2g}$ peak, which is the in-plane phonon vibration mode at 1366.4 cm$^{-1}$, where the thickness was ~ 6 nm as monitored from the AFM topography image and the line profile (**Figure 2c**). Choosing the right h-BN thickness is very important if we are to promote Fowler - Nordheim (FN) tunnelling and as well as protect the ReS$_2$ channel after charge storage in conductive monolayer graphene. **Figure S3 (a, b)** shows that the voltage at which the current start increase rapidly for 7 nm thick h-BN is around 10 V. **Figure S3 (c)** distinguish different region of direct tunnelling and F-N tunnelling with the F-N equation fitting at higher reverse bias region. **Figure S3 (d)** shows the flat band alignment with the values of various parameters of the vertically integrated ReS$_2$/h-BN/graphene system. We found 5-7 nm to be the optimum h-BN thickness for multilevel memory applications. Thickness of h-BN dependent tunnelling behaviour has been demonstrated elsewhere. [25]

## 2.1. Electro-static Memory Operation

The representative output characteristics ($I_d$-$V_d$) for different $V_g$ values (-30 V to +30 V) of a few-layer ReS$_2$/h-BN/FG-graphene FET structure shows a linear $I_d$-$V_d$ relation representing good electrical contacts with low contact resistance at both source and drain under a sweep $V_d$ bias without hysteresis (**Figure 3a**). Linear $I_d$-$V_d$ characteristics under various back-gate-dependent voltages indicate ohmic contact between the ReS$_2$ channel and source/drain electrodes. The representative transfer characteristics ($I_d$-$V_g$; sweep $V_g$) for different $V_d$ values represent the formation of a large hysteresis width for each drain bias (**Figure 3b, Figure S4b**). The device exhibits n-channel charge transport with an electron mobility of around 10-20 cm$^2$ V$^{-1}$ s$^{-1}$ and an ON/OFF current ratio of ~ $10^4$ at $V_d$ = 50 mV. The retention (**Figure 3c, Figure S4c**) of the program/erase state of the memory device was measured with a $V_d$ of 50 mV for different gate bias $V_g$ values. The charges trapped in the graphene layer are maintained without any significant loss. We monitored the program/write state and erase state current over 1000 s with positive (+30 V, 5 s) and negative (-30 V, 5 s) gate voltage pulses, respectively. The logical '0' and '1' outputs are defined with the drain output current of the transistor below and above a current of 500 nA, respectively. Good data retention properties of the NVM devices were achieved in the time range of ~ $10^4$ s (**Figure S5**). Previously thermal assist memory operations have been studied in ReS$_2$, MoS$_2$ based FET devices, where it was found that the intrinsic oxide traps, intrinsic defects/traps in channel material and/or charge trapping and de-trapping between the oxide and p$^+$Si gate are the various reasons for the hysteresis modulation in memory operation. [26, 27] Thermal stability of the retention properties and hysteresis width of the ReS$_2$/h-BN/graphene memory operation are tested (**Figure S6, S7**), which indicates that the memory operation is almost independent of temperature variation. Moreover, ReS$_2$/h-BN FET shows very less temperature dependent transfer characteristics, which further nullify the possibility of thermal assist memory operation of the ReS$_2$/h-BN transistors (**Figure S8**). Bottom layer of h-BN could be assisting to reduce the intrinsic oxide traps and/or charge trapping and detrapping process between the oxide and $p^+$Si gate induce effects on the transistor performance. **Figure S9** and **S10** show the NVM operation of few-layer ReS$_2$ with different thickness of h-BN, where monolayer graphene was used as FG-gate. To gain insight regarding the dynamics of charge carrier trapping and tunnelling and their role in terms of transfer characteristics when producing hysteresis under a sweep bias, we have monitored the transport characteristics while pulsing the gate bias value from a program/write state - `OFF` state (+ 30 V) to an erased state - `ON` state (- 30 V) (**Figure 3d, Figure S4d**). The drain current was recorded, while the gate was repeatedly pulsed between V$_g$ = -30 V and 30 V. When we switched the pulse from a large negative bias to a positive bias, there was a sudden fast rise in the current in the channel followed by a slow decay. Whereas, when the gate pulse was returned to a high negative bias, the current decreased followed by a slow increment. The decay in the ON state during the program is due to electrons being captured in the traps of the bulk defects in the ReS$_2$ or ReS$_2$/h-BN interface states or the defect states of h-BN. On the other hand, the slow current increment in the OFF state is due to the electron emission from the captured states and/or gate field stress induced carriers' leakage (**Figure S11** & **Supplementary Note 1**). [19, 28]

Before we investigated multilevel photoelectric memory operation, we looked at two-level memory operation via electro-static coupling. The sweep transfer characteristics (I$_d$-V$_g$) for different V$_{g,max}$ values represent a large hysteresis window with a high sweep bias operation (**Figure 4a**), which is shown on a logarithmic scale plot in **Figure S12c**. The shift of the threshold voltage in a positive direction corresponds to electron trapping in FG-graphene. The deduced threshold voltage shift (ΔV) as a function of V$_{g,max}$ is summarized in

**Figure 4b**. The amount of charge stored in the FG-graphene as a charge-trap layer can be estimated from the following expression:

$$n = \frac{\Delta V \times C_{FG-CG}}{q} \quad (1)$$

where, q is the electron charge, and $\Delta V$ is the difference between the threshold voltages for the read and erase states. $C_{FG-CG}$ is the capacitance between a floating gate and the control gate and is defined as $C_{FG-CG} = \frac{\varepsilon_0 \varepsilon_{SiO2}}{d_{SiO2}}$, where $\varepsilon_0$ and $\varepsilon_{SiO2}$, respectively, are the vacuum permittivity and relative dielectric constant of the SiO$_2$ layer and $d_{SiO2}$ is its thickness (285 nm). This results in a stored carrier density of $n = \frac{\Delta V \times C_{FG-CG}}{q} = 2.9 \times 10^{12}$ cm$^{-2}$ for $\Delta V = 38$ V (memory window) under a sweep gate bias of ± 30 V (V$_{g,max}$). There may be a small error in the calculated value of the stored carrier's density in FG-graphene caused by the charging/emission process of graphene-SiO$_2$ interface traps states. In agreement with other reports [29, 30, 31], the stored electron density allows a large memory window with strong potential for multi-bit data storage. Here we estimated the electron trapping rate in floating-gate graphene and quantify the tunnel current value by using the following expression:

$$\frac{dn_{trap}}{dt} = \frac{C_{ox}}{q} \frac{dV_g}{dt} \quad (2)$$

Here, $\frac{dV_g}{dt} \sim \frac{\Delta V_s}{\Delta t}$ and $\Delta V_s$ is the gate voltage (10 V) shift on a $\Delta t$ (200 ms) time scale. Thus, we estimate the charge trapping rate to be $\frac{dn_{trap}}{dt} \sim 3.8 \times 10^{12}$ cm$^{-2}$ s$^{-1}$. The charge trapping rate is relatively fast compared with the typical value for a metal-insulator-semiconductor, which is of the order of ~ $10^9$ cm$^{-2}$ s$^{-1}$. A faster charge trapping rate than a typical MOS based memory device is another advantage of the NVM device, and this could be due to the reduced body thickness of the tunnelling layer h-BN in a vertical integrated heterostructure device. Finally, we estimate the tunnel current magnitude using the following formula:

$$\frac{dn_{transfer}}{dt} = \frac{I_{tunnel}}{qA} \quad (3)$$

Where, $I_{tunnel}$, q and A are the tunnel current, electric charge and h-BN cross-sectional area, respectively. As calculated earlier, $\frac{dn_{transfer}}{dt}$ ~ $2.3 \times 10^{15}$ cm$^{-2}$ s$^{-1}$, which provides a tunnel current value of $I_{tunnel}$ ~ $18.4 \times 10^{-12}$ A.

**Figure 4c** shows the dynamic behaviour (logarithmic scale plot in **Figure S12c**) of the NVM device in a single cycle with multiple pulses of low positive and negative gate bias, which includes programming, readout, erasing and readout after erasing processes under a cycle of electrostatic gate pulses as shown in **Figure 4d**. Data reliability tests of NVM devices are measured for up to 1000 cyclic measurements of program/erase operation, which is shown in the **Figure S13** cyclic endurance measurements. The fluctuations in the cyclic tests are quite small up to 1000 cycles operation, suggesting that the programmed data in the memory device is highly reproducible. Moreover, we have studied the tread-off relation between gate pulse width and amplitude of such gate field stressed memory operation (**Figure S14** & **S15**). It is observed that if we reduce the amplitude of the pulse gate voltage below 5 V at fixed width of 10 ms then the memory operation window gets reduce. Similarly, if we fixed the amplitude of pulse gate voltage at 10 V and reduce the pulse width below 5 ms then the memory window decreases. FG-graphene has the advantage that it can store electrons (negative potential) or holes (positive potential) in order to switch the ReS$_2$ channel with a high memory window `OFF` or `ON`. The positive control gate bias of a 10 V pulse with a 10 ms width has the effect of partially storing electrons at the FG-graphene, which turns the operation of the partial write/program state ('0' memory state) of the NVM device. Similarly, a negative gate pulse of (-10 V), which induces the partial storage of holes at FG-graphene, results in `ON` states with a higher ON current magnitude in the ReS$_2$ channel transistor that represents the '1' erase memory state. Next a positive gate pulse of (+10 V) remove the '1' memory states and brings the device current level to an initial `OFF` state corresponding to a low current level, which is program state '0'. The NVM operation and data reliability test of cyclic endurance measurements at +/- 10 V gate bias with 10 ms pulse width from another device is shown in **Figure S9**. The mechanisms of the writing, reading and erasing processes are illustrated in **Figure S16** (Supplementary Figure). A high positive (write/program) and a negative gate bias (erase) with a small drain bias (even at a zero-drain bias), can be used to tune the band alignment to pull electrons from $ReS_2$ and inject electrons into $ReS_2$, respectively. The read operation is performed under a zero gate bias and a low drain bias. Thus, different gate bias strengths can lead to switching to a different current level in a memory device, which can be operated as a multi-level memory device using electrostatic pulses with a range of bias voltages. Successive gate pulses constituting arbitrary electrical stress were used to switch the device between a high (ON) and a low (OFF) current state by applying a positive 10 V (write/program) pulse and a negative 10 V (erase) pulse, respectively (**Figure 4 c, d**). Thus, the programming and erasing processes can be controlled by tuning the polarity of the control back-gate voltage. As a proof of concept, the two current states at a given gate voltage (+/- 10 V) caused by hysteresis can be used as the two logic states of a solid-state memory. The various device parameters and figures-of-merit (FOM) of the multi-bit storage memory device are listed in **Table 1**.

### 2.1. Opto-electrostatic Memory Operation

The memory device has dual functionality; its memory can store both the electrostatic gate pulse state and incident laser pulse information. The laser illumination (532 nm, 4 mW/cm$^2$) can generate both positive and negative photocurrents in the NVM device depending on the applied gate bias for a fixed drain bias of 50 mV (**Figure 5a**). Zero and positive gate biases produce positive photocurrent (PPC) pulses, whereas a negative gate bias produces negative photocurrent (NPC) pulses under a successive ON/OFF cycle of laser pulse illumination. Laser illumination excites electrons from the valence band into the conduction band of $ReS_2$, and simultaneously the application of positive gate pulses for the program/write state induces electron storage in the FG-graphene. Laser irradiation under the programmed state of the NVM device does not allow the memory states to be switched as the charges at the FG-graphene induce holes in the conduction band of the ReS$_2$ channel, which recombines with the photo-generated electrons and makes the channel remain in the OFF state. Therefore, we observe no significant difference in photocurrent ($I_P^{ON}$) generation with laser irradiation under a pulsed positive gate bias (**Figure 5b**). Hence, the laser pulses do not allow us to program the NVM for multi-level program/write states. There was very little photocurrent, $I_p$, generation, and it decreased instantaneously after removal of the laser pulse, however the laser pulse was unable to switch the fully programmed state into a partially programmed state as shown in **Figure 5b**.

Here we describe photo-excited carrier generation and carrier separation under an electrostatic field. A positive gate bias makes it possible to store electrons in FG-graphene, which does not allow the extra photo-generated electron of the ReS$_2$ channel to tunnel through h-BN and the device has a PPC under laser illumination (**Figure 5c**). On the other hand, a zero gate bias has no inward or outward vertical electric field across the ReS$_2$/h-BN/graphene and thus produces a positive photocurrent under laser illumination (**Figure 5d**). The origin of the NPC phenomenon observed under a negative gate pulse and laser pulses in the ReS$_2$/h-BN/FG-graphene vdW heterostructure can be described in terms of the charge transfer between the floating layer graphene and the conduction channel ReS$_2$. Initially we assume the FG-graphene to be hole-doped, as seen in **Figure S2 (b),** and the Fermi energy ($E_F$) shifts below the Dirac point at zero voltage and in a thermodynamic equilibrium condition. Additional positive gate voltage results in a shift of the charge neutrality point of FG-graphene to a more negative voltage. Therefore, photogenerated electrons from ReS$_2$ would not able to occupy already filled energy states in FG-graphene via FN-tunnelling (**Figure 5c**). This results in a net positive photocurrent pulse under a laser pulse. Similarly, when we remove the gate bias (i.e. $V_g$ = 0 V), electrons trapped in the FG-graphene do not allow a photoelectron to tunnel through h-BN. This again creates a positive photocurrent under pulse laser illumination (**Figure 5d**). The opposite scenario will be observed with a negative gate bias, where the gate induces higher hole doping at the FG-graphene. Initially we set the device in the erase state (sweep bias ranging from 40 V to -40 V) then we applied a negative 20 V control gate bias, which made it possible to have a small drain current. The gate bias was set at -20 V, which lowers the $E_F$ level in the higher hole-doped FG-graphene (**Figure 5e**). When there is higher energy laser excitation with the applied drain bias, the generated hot electrons in the $ReS_2$ channel can have sufficient kinetic energy to cross over the h-BN barrier towards the control gate electric field. The photogenerated holes recombine with available free electrons in the $ReS_2$ channel thus reducing the dark current. Therefore, the separation of photogenerated electrons from $ReS_2$ to FG-graphene is not governed by the FN-tunnelling, which results in a sharp decrease in the dark current under laser irradiation. The electrons from the $ReS_2$ will recombine with the holes trapped in the graphene and increase the negativity of the graphene potential. Once the laser is in the `OFF` state, the electrons stored at the FG-graphene move to the $ReS_2$ via FN-tunnelling, which has a slow time response to retain the current state with the initial dark current value. This optical characterization further confirms the carrier transport mechanism and tunnelling phenomena in our NVM device.

Optical erasure is a highly energy efficient and non-contact method designed to achieve multi-level access in the memory device (inset **Figure 6a**). Optical erasure operations with different laser pulse intensities allow us to modulate the logic '1' operation of the memory device. **Figure 6a** shows four different accessible logic '1' states (i.e. $L_1$, $L_2$, $L_3$ and $L_4$) that were realized by varying the laser intensity of the incident laser pulse under an erasing condition, where the laser pulses were programmed and synchronized with the electrostatic gate pulses as shown in **Figure 6b**. Multi-level memory operation under varying the laser intensity are discussed in **Figure S17** & **Supplementary Note 2**. The optical erasure process used to access different erased memory states is not a simple $ReS_2$ channel photo-response process, rather it includes the effect of the heterostructure. The schematic band diagram of the cross-sectional device (**Figure 6b** inset) shows a laser pulse irradiated under an erased condition (holes reached FG-graphene) of the NVM device, which allows different amounts of photogenerated electron tunnelling under different laser intensities. Those electrons recombine with excess holes in FG-graphene and change the current level in the ReS$_2$ channel. By employing higher energy laser excitation with the applied drain bias, the generated hot electrons in the ReS$_2$ channel can have sufficient kinetic energy to cross above the h-BN barrier towards the control gate electric field. However, we do not observe any current level change after using a specific laser intensity of 3 mW/cm$^2$. Now we discuss the dynamic response of the erase operation of the NVM device with electrostatic and opto-electrostatic coupling as shown in the **Figure 6c** and **6d**, respectively.

Previously we have seen two-level i.e. programmed and erased state memory operation under low gate bias pulses via electrostatic coupling. The dynamic response under multiple negative gate pulses at the control gate shows readout erased pulses (ON current) drain current (**Figure 6c**). Now we see that we have a two-level erased state operation with the help of laser irradiation along with electrostatic pulses. We describe the result obtained with pulsed photo illumination cycles under a pulsed gate bias (**Figure 6d**). Binary number digits are used to represent three different memory states as `00` – fully write, `11` – fully erase and `(11')$_{\text{after laser ON}}$` – partially erase state, which are labelled in **Figure 6d** with two different drain current levels, I$_1$ and I$_2$, corresponding to two different erased states, respectively. The device is illuminated by a laser pulse (duration 1 s, 532 nm, intensity 4 mW/cm$^2$) under a zero-gate pulse after the programmed/write operation of the memory device, which results in a noticeable PPC as the photon-excited electrons in the ReS$_2$ channel contribute to the total current. A negative gate bias of 20 V allows the NVM device to readout the drain current corresponding to the fully erase state - `11`. On turning the laser on, the photocurrent (I$_P$) increases sharply and the I$_P^{ON}$ magnitude does not saturate at a fixed value;

instead we observed a slow decrement at a 0 V gate bias. The slow degradation in the current state could be due to the presence of various interfaces (e.g. Au-ReS$_2$; ReS$_2$/h-BN) and/or defect trap states (ReS$_2$; h-BN and ReS$_2$/h-BN), where photogenerated carrier electrons become trapped in defect/interface states during light exposure. This is also observed in this NVM device at V$_g$ = 0 V (**Figure 5a**) and in other reported optoelectronic memory devices. [18] There could be other factors such as defects and trap states at the ReS$_2$, h-BN and ReS$_2$/h-BN interface, which maintains a decrement in the total readout drain current along with the I$_p^{ON}$ magnitude. [19] And turning the laser off does not simply remove the photocurrent component completely, instead the drain current saturates at a new (smaller) magnitude I$_p^{OFF}$, indicating a new readout current state corresponding to a partially erased state - `11`. The optical erasure process is fast at less than 10 ms (limit set in measurement instrument) as the current level immediately reaches a new accessible state 11' after the laser has been turned off. For the fully programmed/write state (`00`), a positive gate pulse (+20 V) can be used as discussed above in **Figures 3** and **4**. Here we discuss the operational mechanism of the ReS$_2$/h-BN/FG-graphene optoelectronic memory. The photo-memory operation is investigated in this 2D heterostructure device by employing the modulation of the conductance of a multilevel ReS$_2$ field-effect transistor via the simultaneous application of laser pulses and electrostatic gate potential pulses. The main mechanism relies on the manipulation of a charging effect at the FG-graphene of the NVM device by the transfer of photogenerated and/or electrostatic induced free carriers of the ReS$_2$ channel through the insulating tunnel layer of h-BN. The offset (time: 2 s) synchronization of the laser pulse (fixed power and frequency) with a gate bias pulse (fixed voltage and frequency) allows the operation of a multi-bit memory storage device. Furthermore, synchronization of the gradual increase in the optical power of the laser pulses (light-controlled) with the gradual increase in the positive gate bias voltage (voltage-controlled) could lead to a more robust multifunctional device along with the ability to realize multibit photo-signalling memory and all-optical logic processing devices.

## 3. Conclusion

In summary, our work demonstrates the non-volatile memory operation of few-layer $ReS_2$/h-BN/FG-graphene-based FET devices. The memory device response can further expand the possibilities of multi-level application via both optical laser excitation and pulse gate bias. Few-layer direct gap $ReS_2$ acted as both a channel material and a light-absorption layer, which allowed us to demonstrate memory operation via opto-electrostatic coupling. Multi-level operation can be achieved depending on the laser intensity and number of pulses for a fixed intensity. We demonstrated multi-level memory operation via 532-nm laser pulse and electrostatic gate pulse coupling. The robustness and stability of the laser-assisted tuneable $ReS_2$/h-BN/FG-graphene memory show the potential for multibit information storage. Our results extend the possibility of multi-bit memory application via opto-electrostatic coupling in 2D heterostructures.

## 4. Experimental Section

*Methods*: Sample Preparation and Measurement Techniques.

Standard dry-transfer techniques [32, 33] were used to prepare a heterostack consisting of $ReS_2$, h-BN and graphene. Initially a $SiO_2$ (285 nm)/Si substrate was used, which was treated with oxygen plasma (100 % $O_2$, 300 W, 20 Pa for 5 minutes) to remove contaminants and to induce a chemically and structurally modified surface layer on the $SiO_2$ with functional silanol groups (Si-OH). [7] Then mechanically exfoliated graphene flakes from HOPG were deposited on the surface of the treated $SiO_2$ at a substrate temperature of 180 $^0$C.

Next an h-BN layer and $ReS_2$ flakes were transferred to the top of the graphene/$SiO_2$ sample using the transfer procedure shown in **Figure S1**, whereby we fabricated a large uniform area of $ReS_2$ flakes and transferred it to the top of the h-BN/graphene/$SiO_2$/Si substrate. A $SiO_2$ (90 nm)/Si substrate was spin-coated (3000 rpm, 60 s) with a water-soluble layer of polyacrylic acid (PAA) and then heated on a hotplate for 5 min at 110°C. The second layer of PMMA-A6 was spin-coated (3000 rpm, 60 s) and then heated on a hotplate at 150°C for 5 min. Bulk $ReS_2$ crystal (HQ Graphene supplier) was mechanically exfoliated on thermal tape (Nitto Denko, model NO319Y-4LSC) just before Au metal deposition. Au (100 nm thick) was sputtered directly onto the exfoliated $ReS_2$/thermal tape using an electron-beam deposition technique. The fresh thermal tape was used to exfoliate the Au/$ReS_2$ flakes from the Au/$ReS_2$ coated thermal tape. Then, Au-mediated [34] exfoliated $ReS_2$ flakes were pasted onto the PMMA/PAA/$SiO_2$/Si substrate at 100°C. The flake/PMMA was detached from the wafer in a water bath and then transferred to a polypropylene carbonate coated PDMS stamp holder. The target flake was aligned with the heterostructure of h-BN/graphene using a microscope. Contact was made, followed by a slow increase in the substrate temperature to 110°C for 5 min. The substrate temperature was maintained, and the PDMS stamp was detached from the sample holder. After the $ReS_2$ flakes had been transferred to the desired junction, the sample was then heated in a vacuum at 130°C for 30 min to enhance the heterojunction connection. Standard electron beam lithography (EBL, ELS-7000, F125 KV) techniques were used to pattern the electrodes on the fabricated heterostructure device. An electron beam sputtering unit was used to deposit Cr/Au (Cr: 5 nm in 0.03 nm/s deposition rate, Au: 50 nm, 0.15 nm/s deposition rate in a high vacuum of $10^{-5}$ Pa).

All electrical measurements were performed in the standard three-probe measurement configuration. The current–voltage (*I–V*), current–time (*I–t*) and all the optoelectrical characteristics (photocurrent-time) of the device were measured using an Agilent 2636A and a semiconductor device analyser (Agilent B1500A) source-measurement unit. The gate pulses and I-t characteristics were recorded with a digital oscilloscope (Tektronix TBS1052B (50 MHz, 2 ch, USB)) and a Keysight (Agilent) 33220A function generator which was synchronized with the source-measurement unit, was used to produce gate pulses. The devices were tested in a high-vacuum chamber ($5 \times 10^{-3}$ Pa) in a Lakeshore probe station at room temperature. The optical coupling with the memory device was achieved using a pulsed-wave laser beam emitted from a diode laser (532 nm, diode-pumped solid-state DPSS laser), which was synchronized with the source-measurement unit. A power meter (Ophir Optics, PD300) was used to measure light intensity. The laser beam irradiated the device directly through the transparent glass window of the Lakeshore vacuum chamber. An atomic force microscope (Olympus/SHIMADZU, Nano search microscope, model OLS3500/SFT-3500, dynamic scanning probe) and a Raman microscope (Nanophoton, model Ramanplus, 532 nm laser, with ×100 - 0.9 N.A. objective lens and 1200 lines/mm grating) were used for the thickness measurement and sample characterisation, respectively.

**Supporting Information**
Supporting Information is available from the Wiley Online Library or from the author.

**Acknowledgements**
This research was supported by the World Premier International Center (WPI) for Materials Nanoarchitectonics (MANA) of the National Institute for Materials Science (NIMS), Tsukuba, Japan with a Grant-in-Aid for Scientific Research (JSPS KAKENHI Grant No./Project/Area No.17F17360). A part of this study was supported by NIMS Nanofabrication Platform and NIMS Molecule & Material Synthesis Platform in Nanotechnology Platform Project sponsored by the Ministry of Education, Culture, Sports, Science, and Technology (MEXT), Japan.

**Keywords**
Two-dimensional materials, $ReS_2$, Graphene, Heterostructures, Memory, Photoelectric Memory, Multibit

Received: ((will be filled in by the editorial staff))
Revised: ((will be filled in by the editorial staff))
Published online: ((will be filled in by the editorial staff))

**References**

[1] S. Bertolazzi, P. Bondavalli, S. Roche, T. San, S. Y. Choi, L. Colombo, F. Bonaccorso, P. Samorì, *Adv. Mater.* **2019**, *31*, 1–35.

[2] Q. H. Wang, K. Kalantar-Zadeh, A. Kis, J. N. Coleman, M. S. Strano, *Nature Nanotechnology*. **2012**, 7, 699–712.

[3] K. S. Novoselov, A. Mishchenko, A. Carvalho, A. H. Castro Neto, *Science* **2016**, *353*, 9439.

[4] G. H. Lee, Y. J. Yu, C. Lee, C. Dean, K. L. Shepard, P. Kim, J. Hone, *Appl. Phys. Lett.* **2011**, *99*, 243114.

[5] L. Britnell, R. V. Gorbachev, R. Jalil, B. D. Belle, F. Schedin, A. Mishchenko, T. Georgiou, M. I. Katsnelson, L. Eaves, S. V. Morozov, N. M. R. Peres, J. Leist, A. K. Geim, K. S. Novoselov, A. L. Ponomarenko, *Science* **2012**, *335*, 947–950.

[6] S. H. Kim, S.-G. Yi, M. U. Park, C. Lee, M. Kim, K.-H. Yoo, *ACS Appl. Mater. Interfaces* **2019**, *11*, 25306–25312.

[7] J. Lee, S. Pak, Y. W. Lee, Y. Cho, J. Hong, P. Giraud, H. S. Shin, S. M. Morris, J. I. Sohn, S. N. Cha, J. M. Kim, *Nat. Commun.* **2017**, *8*, 1–8.

[8] S. Wang, C. He, J. Tang, X. Lu, C. Shen, H. Yu, L. Du, J. Li, R. Yang, D. Shi, G. Zhang, *Adv. Electron. Mater.* **2019**, *5*, 1800726.

[9] A. C. Gadelha, A. R. Cadore, K. Watanabe, T. Taniguchi, A. M. De Paula, L. M. Malard, R. G. Lacerda, L. C. Campos, *2D Mater.* **2019**, *6*, 025036.

[10] C. Liu, X. Chen, J. Li, J. Wang, J. Liu, D. W. Zhang, W. Hu, P. Zhou, *Adv. Mater.* **2019**, 31, 1808035.

[11] S. Bertolazzi, D. Krasnozhon, A. Kis, *ACS Nano* **2013**, *7*, 3246–3252.

[12] Q. Wang, Y. Wen, K. Cai, R. Cheng, L. Yin, Y. Zhang, J. Li, Z. Wang, F. Wang, F. Wang, T. A. Shifa, C. Jiang, H. Yang, J. He, *Sci. Adv.* **2018**, *4*, 1.

[13] K. Roy, M. Padmanabhan, S. Goswami, T. P. Sai, G. Ramalingam, S. Raghavan, A. Ghosh, *Nat. Nanotechnol.* **2013**, *8*, 826–830.

[14] B. Hwang, J. S. Lee, *Adv. Elect. Mater.* **2019,** 5, 1800519.

[15] J. Lee, S. Pak, Y. W. Lee, Y. Cho, J. Hong, P. Giraud, H. S. Shin, S. M. Morris, J. I. Sohn, S. N. Cha, J. M. Kim, *Nat. Commun.* **2017**, *8*, 14734.

[16] F. Zhou, J. Chen, X. Tao, X. Wang, Y. Chai, *Research* **2019**, 2019, Article ID 9490413.

[17] D. Lee, E. Hwang, Y. Lee, Y. Choi, J. S. Kim, S. Lee, J. H. Cho, *Adv. Mater.* **2016**, *28*, 9196–9202.

[18] S. Lei, F. Wen, B. Li, Q. Wang, Y. Huang, Y. Gong, Y. He, P. Dong, J. Bellah, A. George, L. Ge, J. Lou, N. J. Halas, R. Vajtai, P. M. Ajayan, *Nano Lett.* **2015**, *15*, 259.

[19] D. Xiang, T. Liu, J. Xu, J. Y. Tan, Z. Hu, B. Lei, Y. Zheng, J. Wu, A. H. C. Neto, L. Liu, W. Chen, *Nat. Commun.* **2018**, *9*, 1.

[20] S. Tongay, H. Sahin, C. Ko, A. Luce, W. Fan, K. Liu, J. Zhou, Y. S. Huang, C. H. Ho, J. Yan, D. F. Ogletree, S. Aloni, J. Ji, S. Li, J. Li, F.M. Peeters, J. Wu, *Nat. Commun.* **2014**, *5*, 3252.

[21] K. Thakar, B. Mukherjee, S. Grover, N. Kaushik, M. Deshmukh, S. Lodha, *ACS Appl. Mater. Interfaces* **2018**, *10*, 36512–36522.

[22] B. Mukherjee, A. Zulkefli, R. Hayakawa, Y. Wakayama, S. Nakaharai, *ACS Photonics* **2019**. 6, 2277-2286.

[23] M. Rahman, K. Davey, S. Z. Qiao, *Adv. Funct. Mater.* **2017**, *27, 1606129*.

[24] X. Wang, X. Li, L. Zhang, Y. Yoon, P. K. Weber, H. Wang, J. Guo, H. Dai, *Science* **2009**, *324*, 768.

[25] L. Britnell, R. V. Gorbachev, R. Jalil, B. D. Belle, F. Schedin, M. I. Katsnelson, L. Eaves, S. V. Morozov, A. S. Mayorov, N. M. R. Peres, A. H. Castro Neto, J. Leist, A. K. Geim, L. A. Ponomarenko, K. S. Novoselov, *Nano Lett.* **2012**, 12, 1707−1710.

[26] N. Goyal, D. M. A. Mackenzie, V. Panchal, H. Jawa, O. Kazakova, D. H. Petersen, S. Lodha, *Appl. Phys. Lett.* **2020**, 116, 052104.

[27] N. Kaushik, D. M. Mackenzie, K. Thakar, N. Goyal, B. Mukherjee, P. Boggild, D. H. Petersen, and S. Lodha, *npj 2D Mater. Appl.* **2017**, 1, 34.

[28] I. Amit, T. J. Octon, N. J. Townsend, F. Reale, C. D. Wright, C. Mattevi, M. F. Craciun, S. Russo, *Adv. Mater.* **2017**, *29, 1605598*.

[29] M. Sup Choi, G. H. Lee, Y. J. Yu, D. Y. Lee, S. Hwan Lee, P. Kim, J. Hone, W. J. Yoo, *Nat. Commun.* **2013**, *4, 1624*.

[30] A. Di Bartolomeo, L. Genovese, F. Giubileo, L. Iemmo, G. Luongo, T. Foller, M. Schleberger, *2D Mater.* **2017**, *5*, 015014.

[31] Y. Wang, E. Liu, A. Gao, T. Cao, M. Long, C. Pan, L. Zhang, J. Zeng, C. Wang, W. Hu, S. J. Liang, F. Miao, *ACS Nano* **2018**, *12*, 9513.

[32] F. Pizzocchero, L. Gammelgaard, B. S. Jessen, J. M. Caridad, L. Wang, J. Hone, P. Bøggild, T. J. Booth, *Nat. Commun.* **2016**, *7, 11894*.

[33] T. Iwasaki, K. Endo, E. Watanabe, D. Tsuya, Y. Morita, S. Nakaharai, Y. Noguchi, Y. Wakayama, K. Watanabe, T. Taniguchi, S. Moriyama, *ACS Appl. Mater. Interfaces* **2020**, 12, 8533.

[34] S. B. Desai, S. R. Madhvapathy, M. Amani, D. Kiriya, M. Hettick, M. Tosun, Y. Zhou, M. Dubey, J. W. Ager, D. Chrzan, A. Javey, *Adv. Mater.* **2016**, *28*, 4053.

**List of Figures**

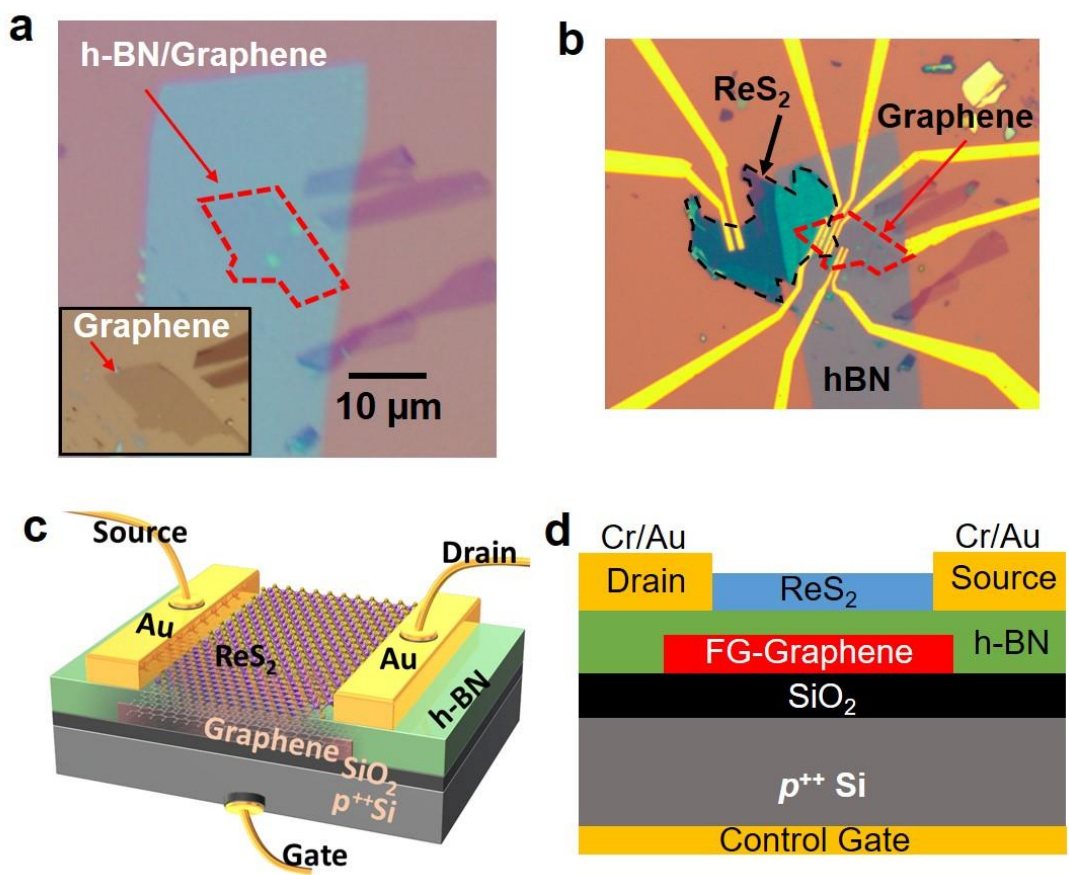

**Figure 1.** (a) Optical image of fabricated heterostructures of h-BN/monolayer graphene. Inset shows monolayer graphene on a $SiO_2$/Si substrate, which was used to prepare the heterostructure device. (b) Optical image of a fabricated multilayer $ReS_2$/h-BN/FG-graphene heterostructure device with electrical contacts. (c, d) Schematic diagrams of 3D and cross-sectional views of the designed device, respectively.

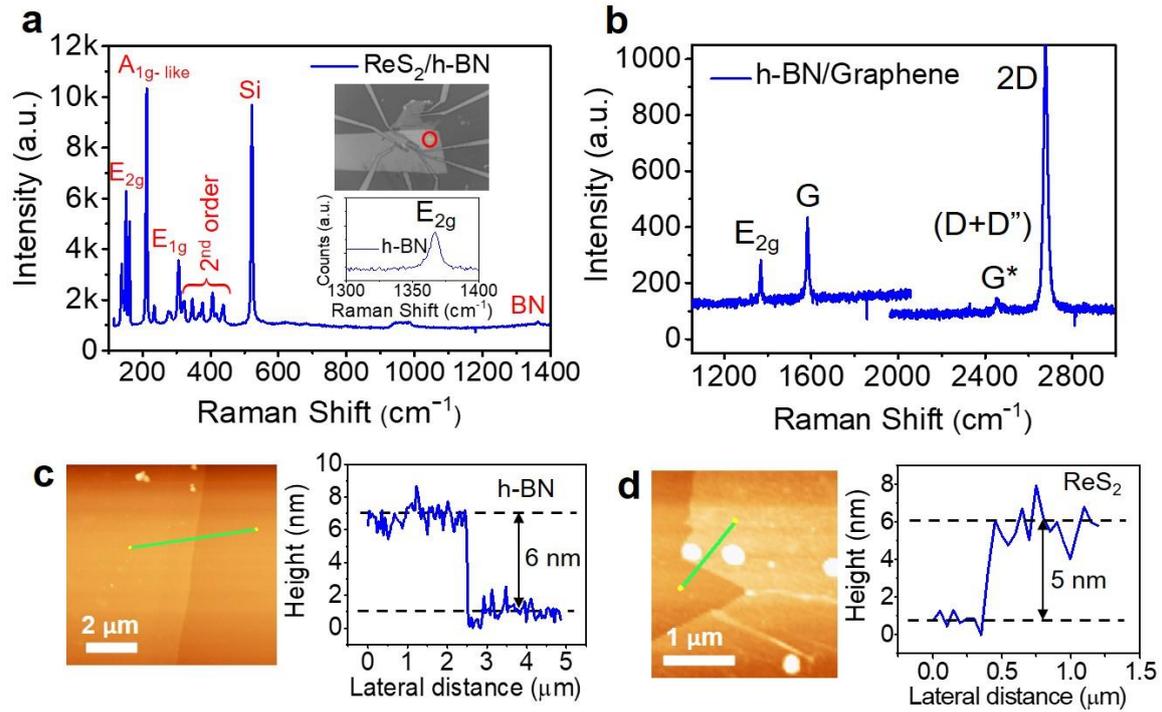

**Figure 2.** (a, b) Raman spectra of the $ReS_2$/h-BN heterostructure and h-BN/graphene/$SiO_2$, respectively. The inset in (a) shows a red circular region in the device optical image from where the Raman spectrum was collected and a close-up of the region corresponding to the h-BN Raman signal. (c, d) Surface topography AFM image and line profile of the thickness across the green line of the h-BN and $ReS_2$ flakes, respectively.

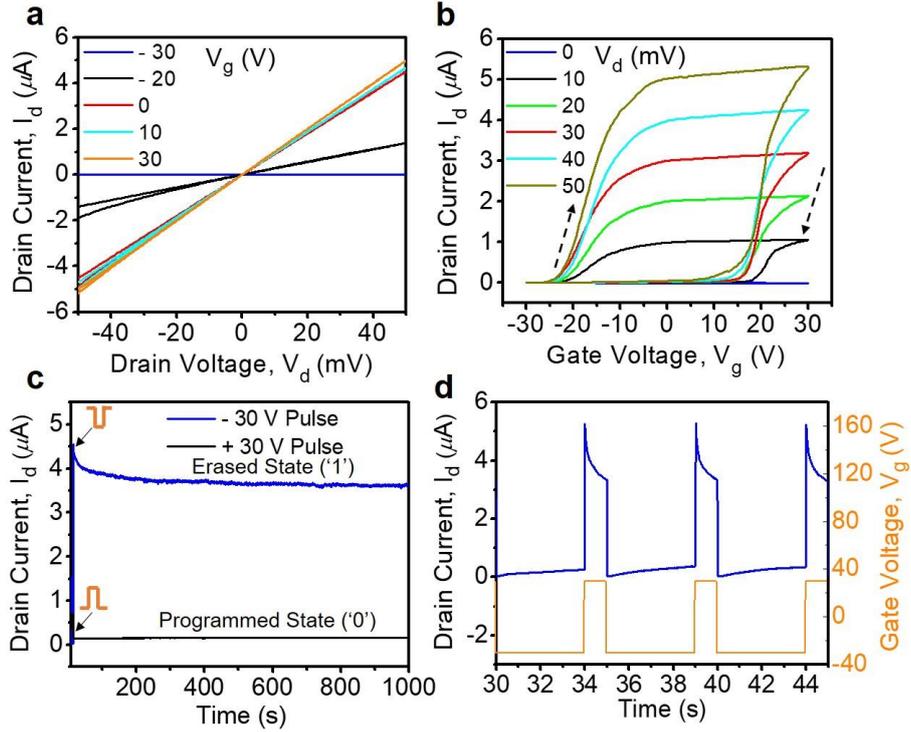

**Figure 3.** Electrical characteristics of a memory device: (a) Output characteristics curves: sweep drain current ($I_d$) versus $V_d$ for different control gate biases $V_g$. (b) Sweep $I_d$ versus $V_g$ for different $V_d$ values, which shows a large memory window. (c) Retention time characteristic of $I_d$ in the ON and OFF states. Each state was read at $V_g = 0$ V, $V_d = 50$ mV after being programmed (erased) by one pulse voltage of $+30$ V ($-30$ V) and a width of 5 s on the control gate. (d) Dynamic switching behaviour of the ON and OFF states induced by applying alternating $V_g$ pulses ($\pm 30$ V, 1 s) with a time interval of 5 s.

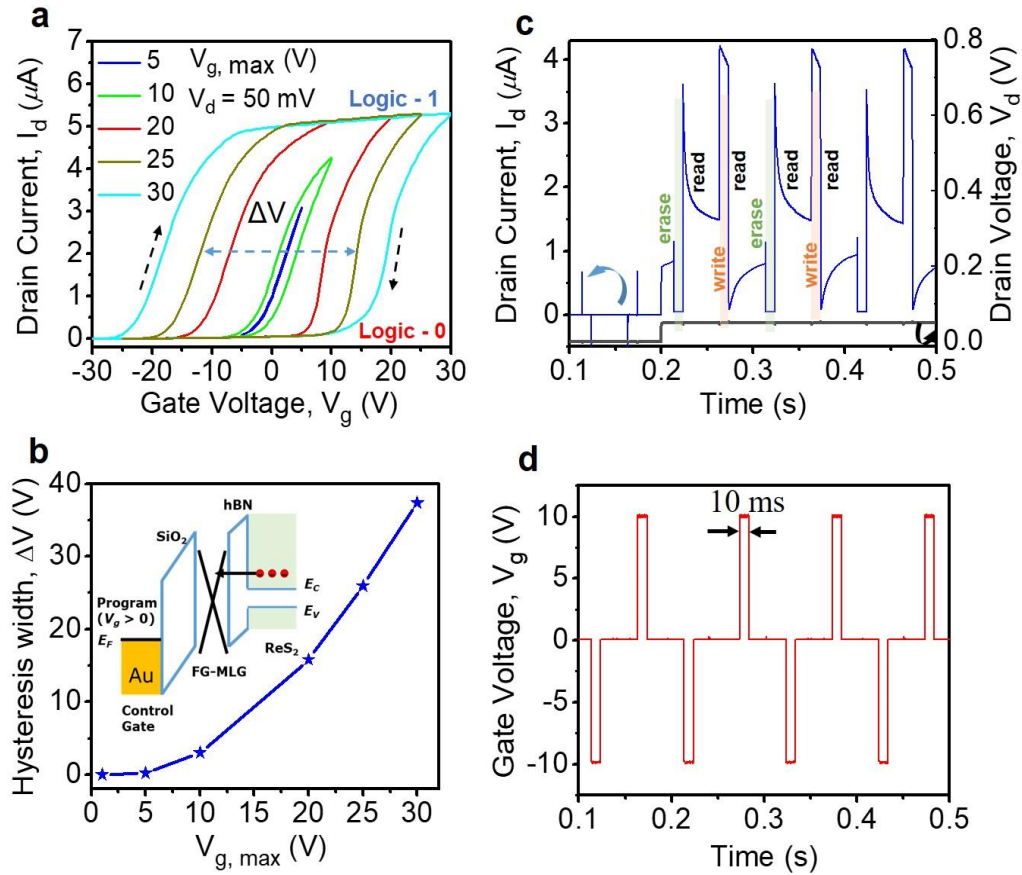

**Figure 4.** Multi-level access via electrostatic coupling: (a) Sweep drain current ($I_d$) versus gate bias ($V_g$) for different gate voltage maxima ($V_{g,max}$), which exhibits a large memory window. A large memory window enables us to realize multi-level operation. (b) Hysteresis width (i.e. memory window) versus $V_{g,max}$. The inset shows the band alignment across the heterostructure of Au/SiO$_2$/graphene/h-BN/ReS$_2$ under a positive gate bias. Successive program/erase operations using an arbitrary gate pulse of +/- 10 V with a pulse width of 10 ms: (c) $I_d$-t response of the NVM device under a drain bias condition of 0 V for 0.2 s and 0.05 V for the next 0.3s, when using an arbitrary gate pulse of +/- 10 V with a pulse width of 10 ms as shown in (d).

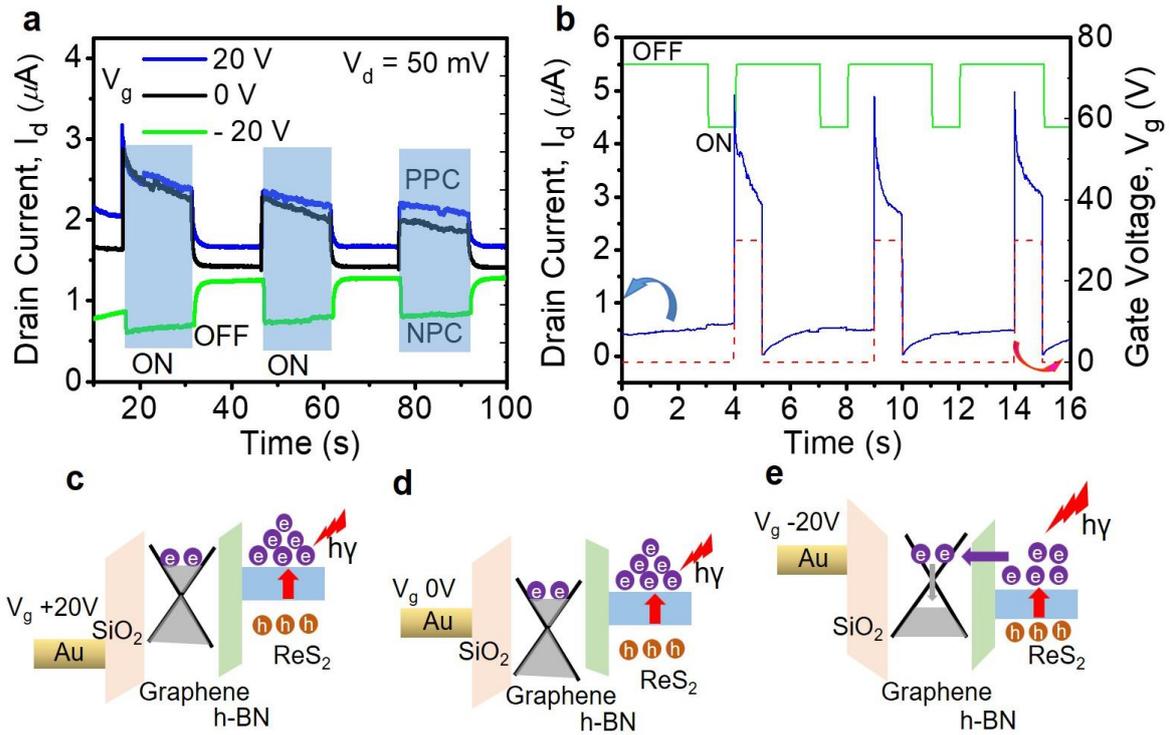

**Figure 5.** Photocurrent generation mechanism in the NVM device with FG-graphene: (a) $I_d$-t response at a fixed $V_d$ of 50 mV for different $V_g$ values under multiple ON, OFF states of the laser (4 mW/cm$^2$, 532 nm), where the laser ON state is shown by blue shading. (b) Dynamic response ($I_d$-t) of the device under both gate pulses (+ 30 V) as shown by the dashed red line, and the incident laser pulses (4 mW/cm$^2$, 532 nm) shown by the green line were used. (c, d, e) Band alignment and positive photocurrent (PPC), zero gate bias and negative photocurrent (NPC) mechanisms under positive, zero and negative gate bias conditions, respectively.

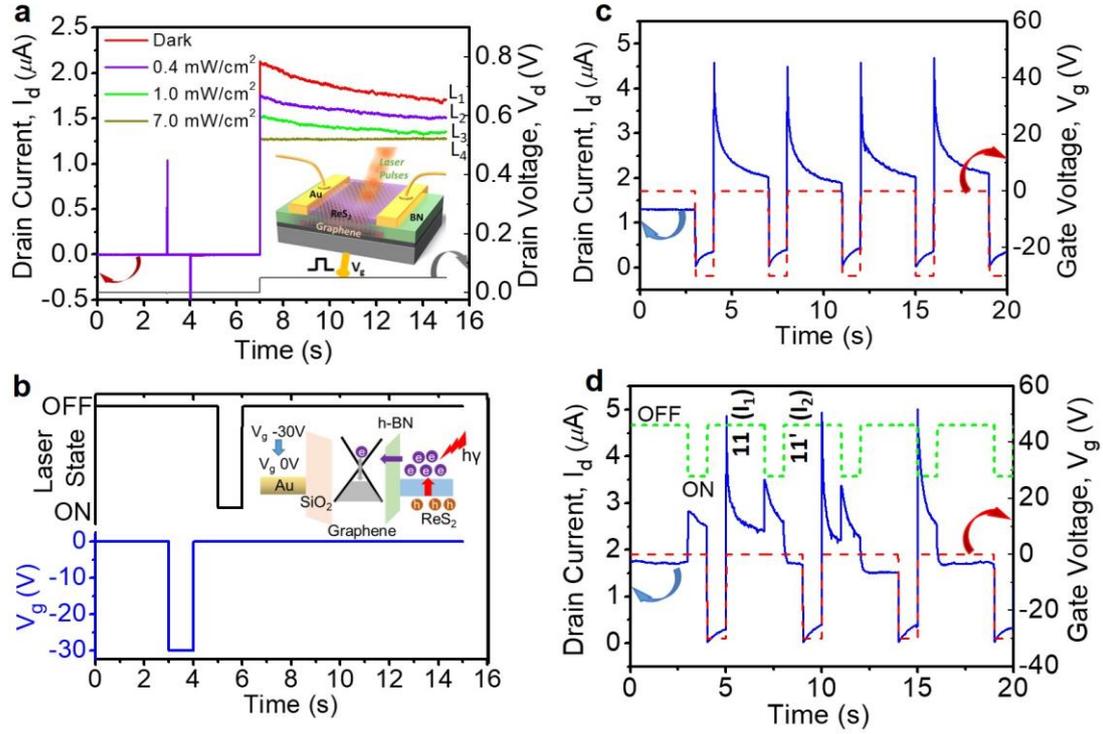

**Figure 6.** Multi-level memory operation via opto-electrostatic coupling: (a) Modulated logic state '1' operation via opto-electrostatic coupling under a pulsed green laser (532 nm) and a pulsed gate bias as shown in (b). Insets show a schematic of the device under optical pulse irradiation and band alignment and the carrier separation mechanism during an optical erasing operation. (c) Multiple electro-static erasure and readout after an erasing operation under a 5s gate pulse. (d) $I_d$-t response of the device to demonstrate multi-level drain current states, where both gate pulses as shown by the red dashed line and incident laser pulses (4 mW/cm$^2$, 532 nm) shown by the green dashed line were used.

**ToC Figure**

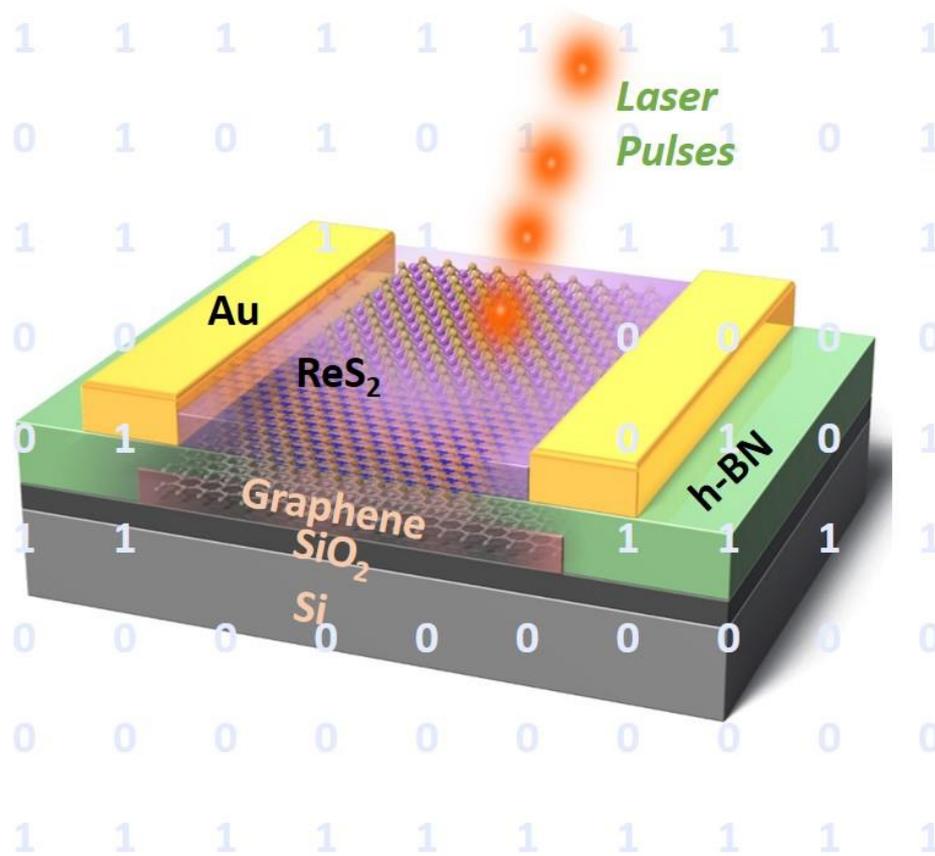

2D heterostructure based non-volatile memory (NVM) with a large hysteresis window to achieve multibit data storage. Demonstration of dual functionality via both electrostatic gate pulses and incident laser pulses. Laser with low intensity communicated multibit storage photoelectronic memory operation at a low operation bias of 50 mV with a good retention time and endurance. Laser assisted fast (10 ms) optical erasure operation using a green laser pulses with a low laser intensity of 1- 4 mW/cm$^2$.

Supporting Information

# Laser-Assisted Multilevel Non-Volatile Memory Device Based on 2D van-der-Waals Few-layer-ReS$_2$/h-BN/Graphene Heterostructures

*Bablu Mukherjee\*, Amir Zulkefli, Kenji Watanabe, Takashi Taniguchi, Yutaka Wakayama, Shu Nakaharai*\**

Dr. B. Mukherjee, A. Zulkefli, Prof. Y. Wakayama, Dr. S. Nakaharai

International Center for Materials Nanoarchitectonics (MANA),

\* Corresponding authors. Email: MUKHERJEE.Bablu@nims.go.jp (B. Mukherjee);

NAKAHARAI.Shu@nims.go.jp (S. Nakaharai)

A. Zulkefli, Prof. Y. Wakayama

Department of Chemistry and Biochemistry, Faculty of Engineering, Kyushu University,

Dr. K. Watanabe, Dr. T. Taniguchi

Research Center for Functional Materials,

National Institute for Materials Science (NIMS), 1-1 Namiki, Tsukuba, Ibaraki 305-0044, Japan.

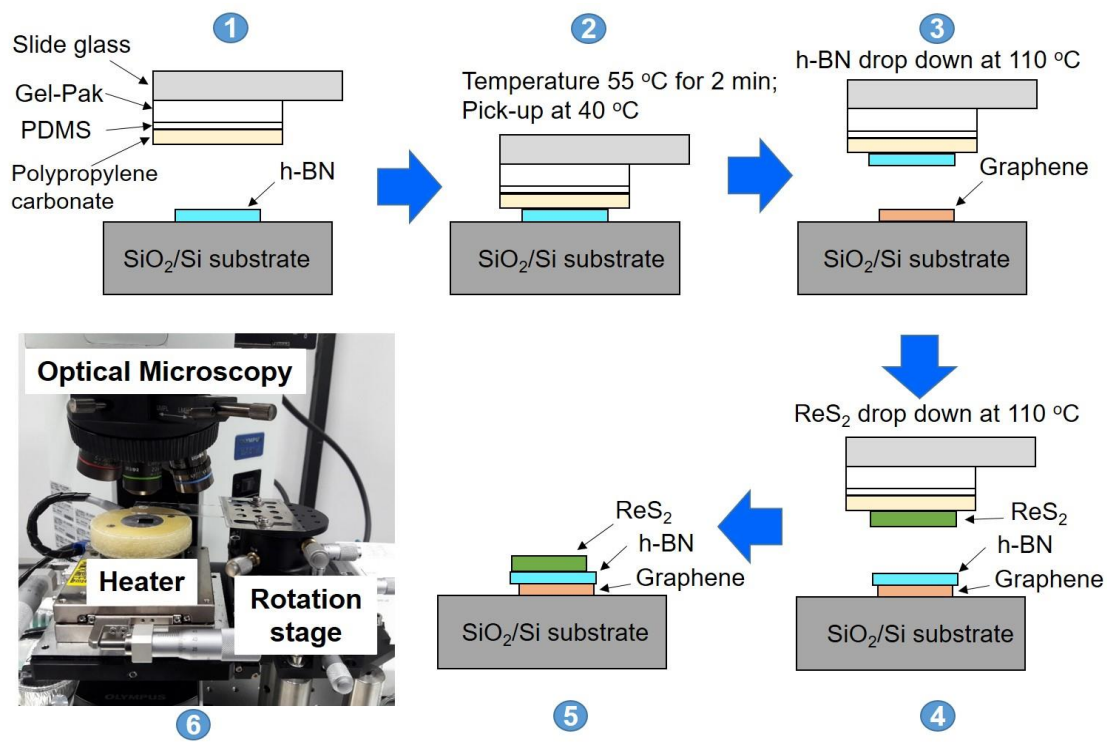

**Figure S1.** Step (1-5): Schematic illustration of the layers transfer techniques via dry transfer method using the optical microscope as shown in step (6). Step (1-2): align the polypropylenecarbonate coated PDMS stamp to pick-up h-BN flake at $40^0$C. Step (3-4): transfer h-BN and ReS$_2$ flake on top of the pre-fabricated graphene/SiO$_2$/Si sample at $110^0$C.

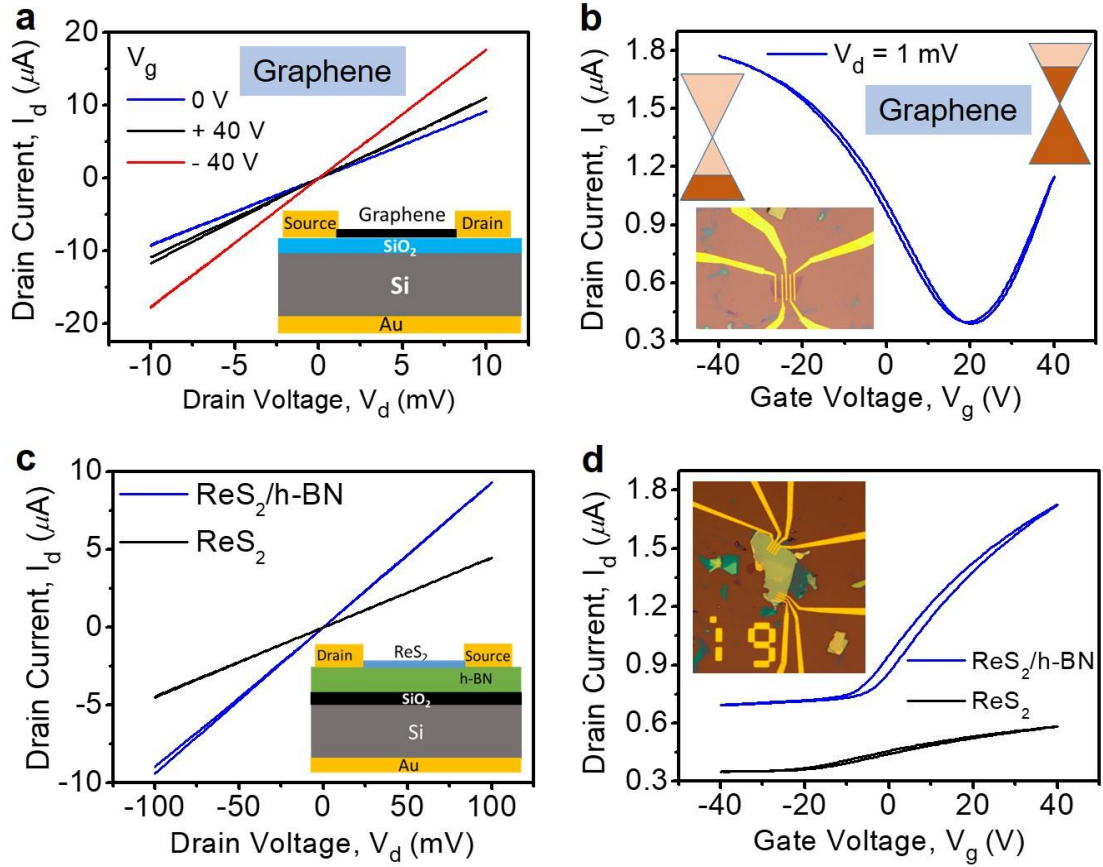

**Figure S2.** (a,b) The output characteristics ($I_d$-$V_d$) and transfer characteristics of the graphene FET. Insets are a schematic and optical image of monolayer graphene FET device. (c,d) $I_d$-$V_d$ at fixed zero gate bias and $I_d$-$V_g$ at fixed 0.01 $V_d$ of $ReS_2$ and $ReS_2$/h-BN FETs. Insets are a schematic and optical image of $ReS_2$/h-BN FET device. There was no significant large hysteresis in the electrical characteristics of graphene/$SiO_2$ and $ReS_2$/h-BN FET devices so the charge trapping phenomena was not at the interfaces among the layers instead it happened in the graphene layer.

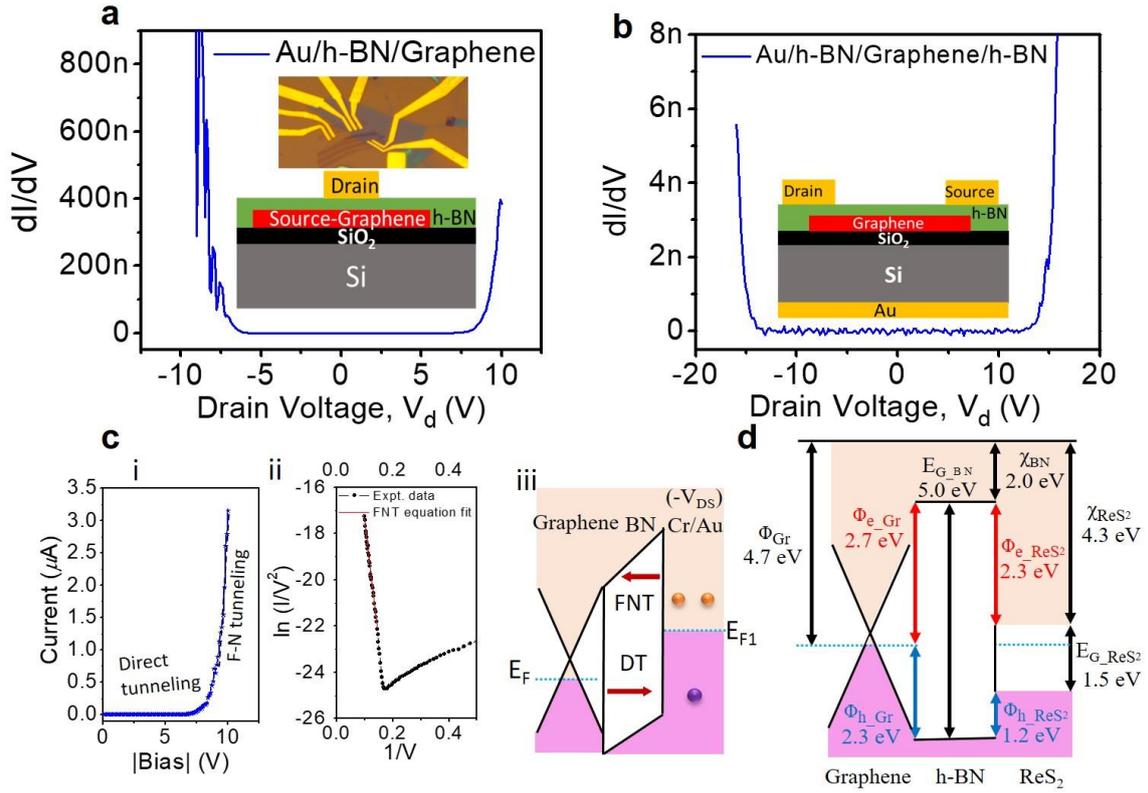

**Figure S3.** (a) Differential conductance dI/dV with bias of Au(Cr)/h-BN/graphene and Au(Cr)/h-BN/FG-graphene/Au(Cr) devices. Insets show a schematic and optical image of the device. (c) Analysis of tunneling behavior in Au(Cr)/h-BN/graphene device as shown in (a): (i) the measured tunneling current as a function of the applied voltage for the reverse bias direction, (ii) $\ln(I/V^2)$ versus $(1/V)$ plot of (i) with F-N tunneling equation fit at higher bias region, and (iii) energy band diagram of Au(Cr)/h-BN/graphene at negative drain voltage at Cr/Au terminal. (d) Flat energy band diagram of vertically integrated $ReS_2$/h-BN/graphene memories, where χ, φ, and $E_g$ represent the electron affinity, work function, and band gap, respectively. [s1]

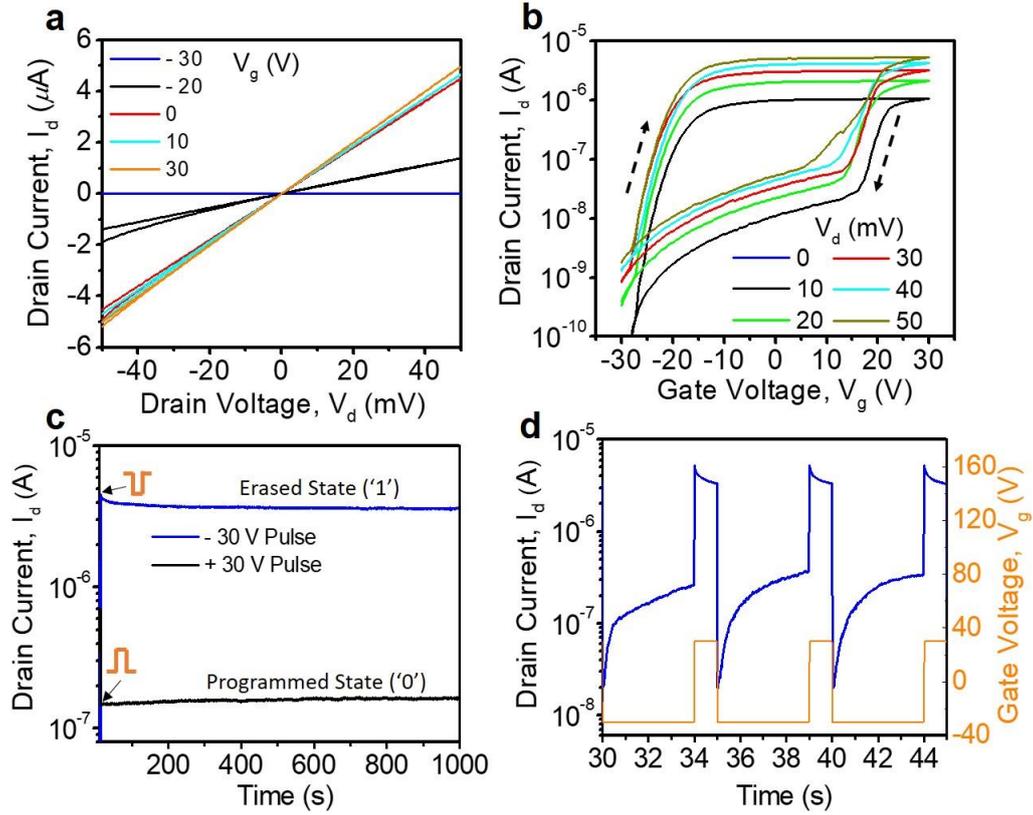

**Figure S4**. Replot of **Figure 3**: Electrical characteristics of a memory device: (a) Output characteristics curves: sweep drain current ($I_d$) versus $V_d$ for different control gate biases $V_g$. (b) Sweep $I_d$ versus $V_g$ for different $V_d$ values on a logarithmic scale, which shows a large memory window. (c) Retention time characteristic of $I_d$ in the ON and OFF states on a logarithmic scale. Each state was read at $V_g = 0$ V, $V_d = 50$ mV after being programmed (erased) by one pulse voltage of + 30 V (−30 V) and a width of 5 s on the control gate. (d) Dynamic switching behaviour of the ON and OFF states induced by applying alternating $V_g$ pulses (±30 V, 1 s) with a time interval of 5 s on a logarithmic scale.

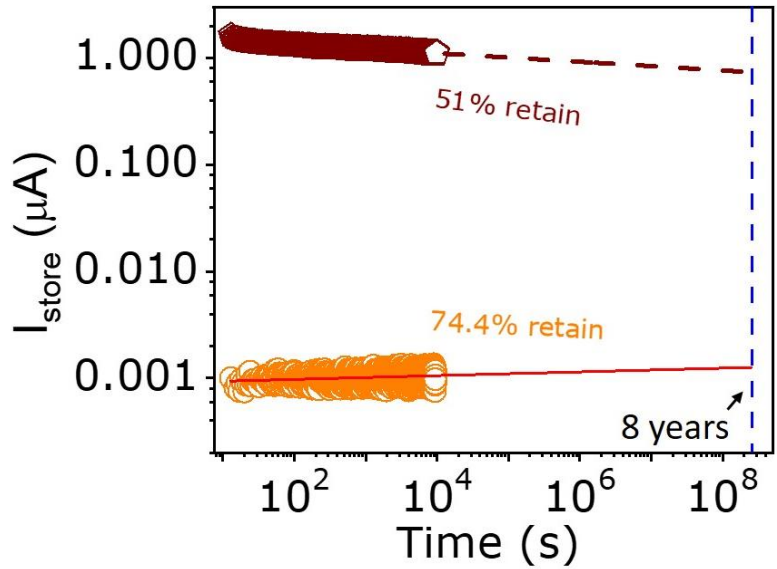

**Figure S5:** Data reliability tests for ReS$_2$/h-BN/graphene memory. Data retention: the storage currents after each programming nearly remain in the time range of ~10$^4$ s, which indicates the good data retention property of the memory device. 8 years' linear extrapolation of the storage currents without electrical and optical pulses after programming and erasing operation. Nearly half the storage information can be retained after 8 years. Drain bias of 0.1 V and pulse gate bias of + 30 V and -30 V of 1s pulse width were used for program and erase states, respectively.

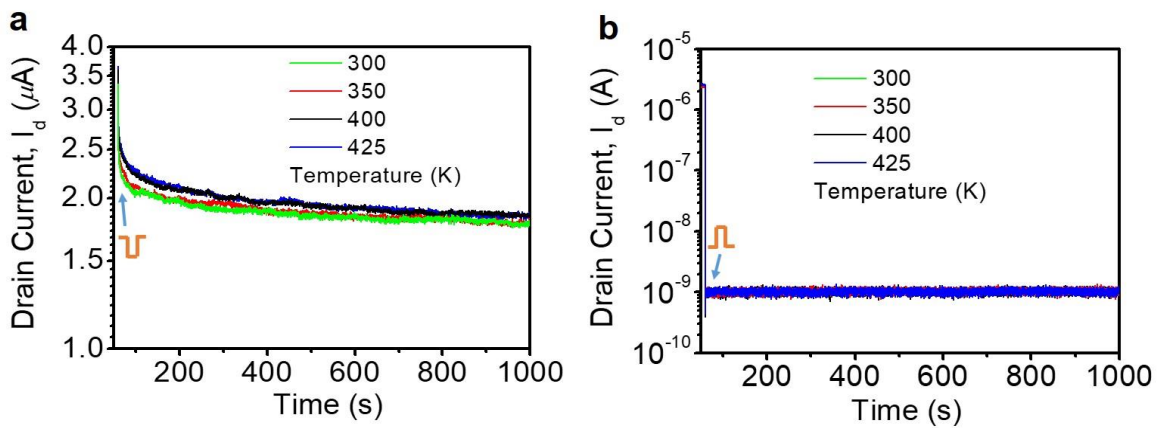

**Figure S6:** The thermal stability of (a) erase states corresponding to high current level (device ON) and (b) program states corresponding to low current level (device OFF). Drain bias of 0.2 V and pulse gate bias of + 30 V and -30 V of 2s pulse width were used for program and erase states, respectively.

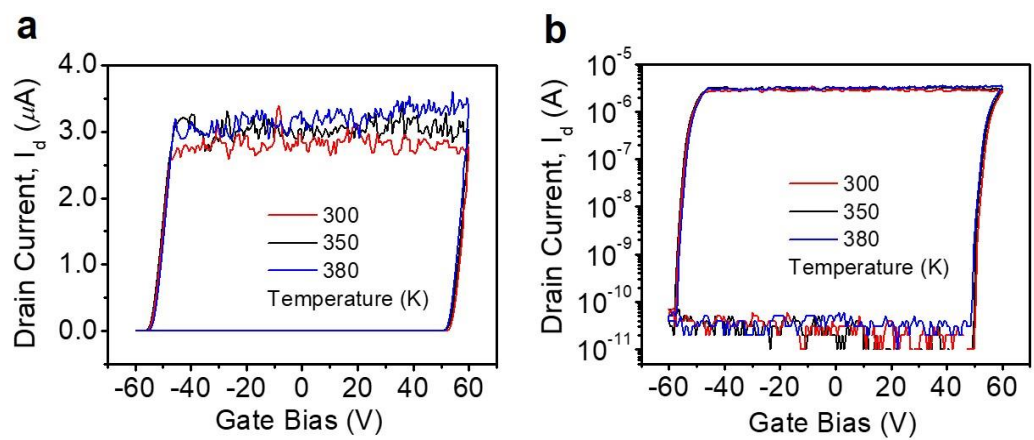

**Figure S7:** The thermal stability test of the memory operation for three different temperatures above room temperature: (a, b) sweep drain current ($I_d$) versus gate bias ($V_g$) in normal scale and logarithmic scale, respectively. There is almost no change in the hysteresis width with temperature.

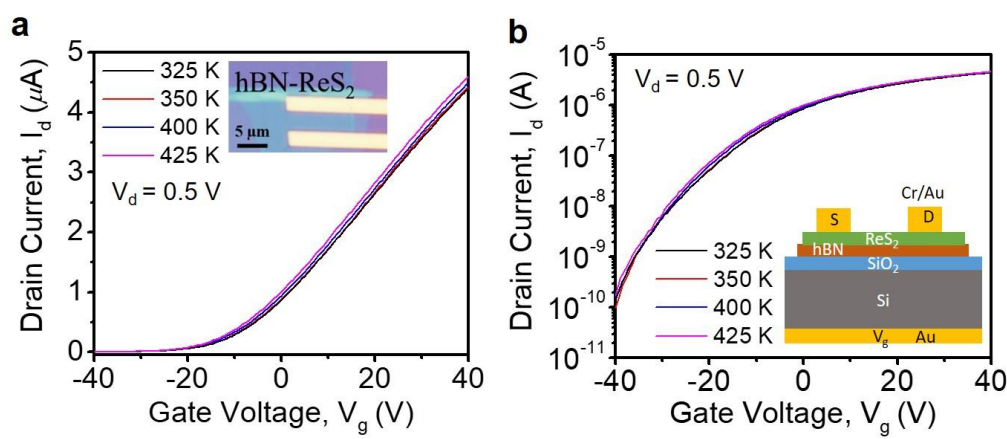

**Figure S8:** The thermal stability of transfer characteristics of $ReS_2$/h-BN FET: (a) $I_d$ versus $V_g$ plot of the transistor for different temperatures above the room temperature. Inset shows the optical image of the tested device. (b) Semilog plot of the transfer characteristics as shown in (a). Inset shows the schematic of the device.

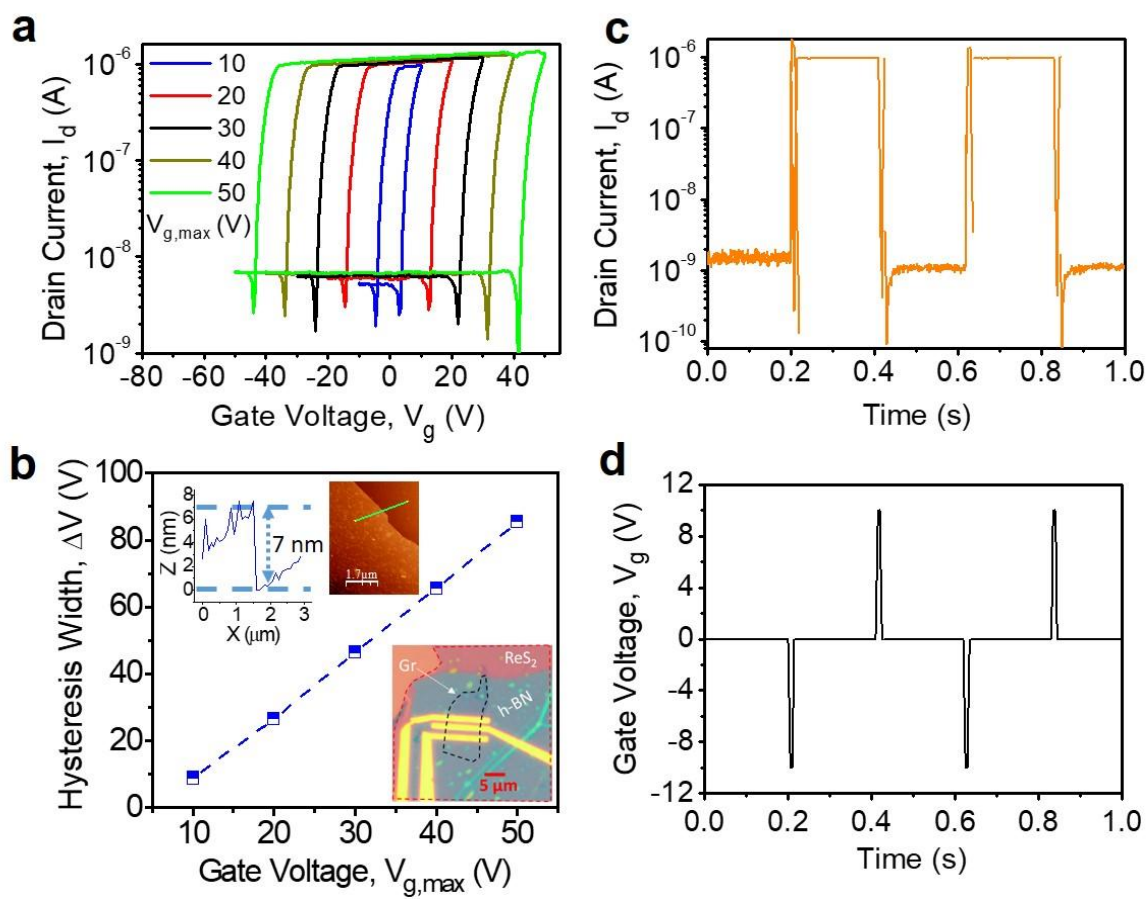

**Figure S9:** Electrical characteristics of a memory device with h-BN thickness of 7 nm. (a) Sweep drain current ($I_d$) versus gate bias ($V_g$) for different gate voltage maxima ($V_{g,max}$) on a logarithmic scale. (b) Hysteresis width (i.e. memory window) versus $V_{g,max}$. The insets show the optical image of the tested device, AFM image and line profile for the thickness measurement of the h-BN layer. (c) $I_d$-t response on a logarithmic scale under a drain bias condition of 0.1 V and gate pulse of +/- 10 V with a pulse width of 10 ms as shown in (d).

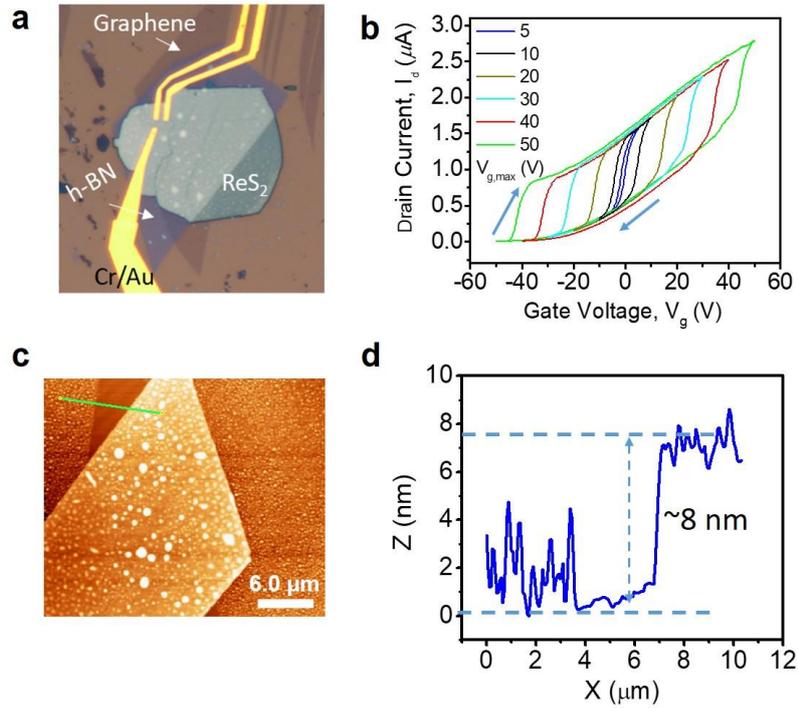

**Figure S10:** Optical image of fabricated multilayered ReS$_2$/h-BN/graphene heterostructure based NVM device. In this device, ~9 nm thick h-BN and monolayer graphene were used. (b) Sweep I$_d$-V$_g$ for different V$_{g,maximum}$ represent the presence of large hysteresis width. (c, d) AFM image of h-BN layer and line profile across the green line in (c) determines the thickness.

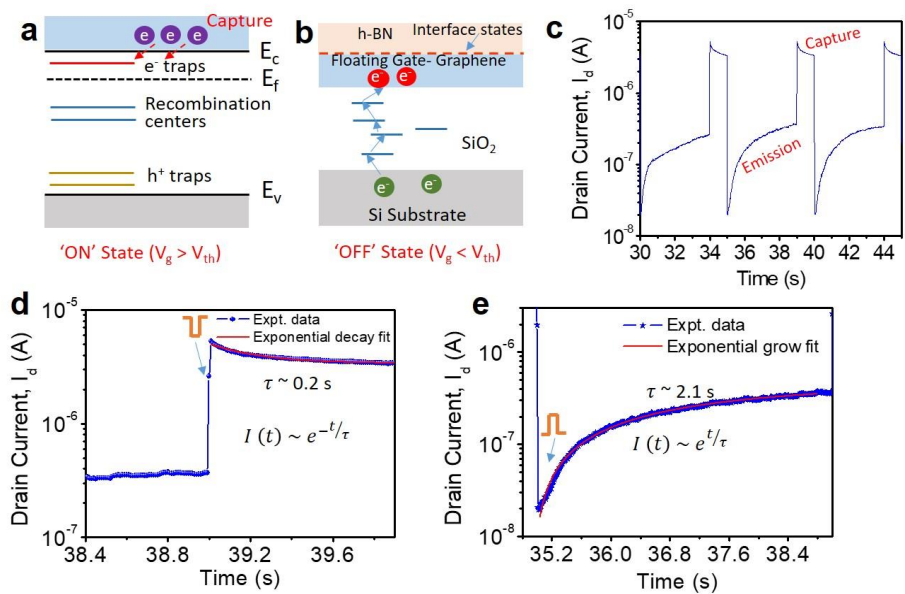

**Figure S11**: (a) Schematic energy-band diagram of n-type ReS$_2$ under positive electric field. Various traps states are labelled with the possibilities of electron capturing process at trap states near to conduction band. (b) Schematic representation to show various possibly of emission process during the negative electric field. (c) Replot of figure 3d: dynamic switching behavior of the ON and OFF states in logarithmic scale induced by applying alternating V$_g$ pulses (±30 V, 1 s) with a time interval of 5 s. (d, e) Zoom-in I$_d$-t characteristics shows decay process during capture and grow process during emission along with exponential decay and grow fits, respectively.

**Supplementary Note 1: Capturing and emission process during pulse gate voltage tuned memory operation**

It has been shown experimentally $^{s2}$ that conventional ReS$_2$/SiO$_2$ FET device contains large amount of various localized defects states, which can act as various recombination centres and/or traps states under laser illumination and/or gate field stress and/or thermally activated. 'S' vacancies produce donor like trap states ($10^{10}$ cm$^{-2}$) below CB, which can be contributing in the capture process. There might exist large density of deep traps ($10^{13}$ cm$^{-2}$) in as-prepared ReS$_2$ samples and relatively less density of acceptor like trap states, which capture/emit holes as shown in schematic energy-band diagram **Figure S11a**. However, the temperature dependent $I_d$-$t$ characteristics (**Figure S6**) and laser intensity dependent characteristics $I_d$-$t$ (**Figure S17**) have confirmed that the decay and rise in $I_d$-$t$ (**Figure S11c**) depends on the applied electric field stress.

Fast decay (coefficient 0.2 s, **Figure S11d**) and comparable slow rise (coefficient 2.1 s, **Figure S11e**) from the exponential fit suggest that the gate pulse induced trapping/detrapping process at various interfaces. During ON state, the capture process is quite fast and spontaneous, which is independent of thermally activation (**Figure S6**) and it could be due to electrons traps in the shallow defect levels present in ReS$_2$ channel and/or defects/traps at ReS$_2$/h-BN interface. On the other side during off state, the emission process is slow, which could be due to the gate field stress induced leakage through trap-assisted hopping conduction of electrons (**Figure S11b**) and/or due to carriers are emitted from the traps states of ReS$_2$, **Figure S11a**.

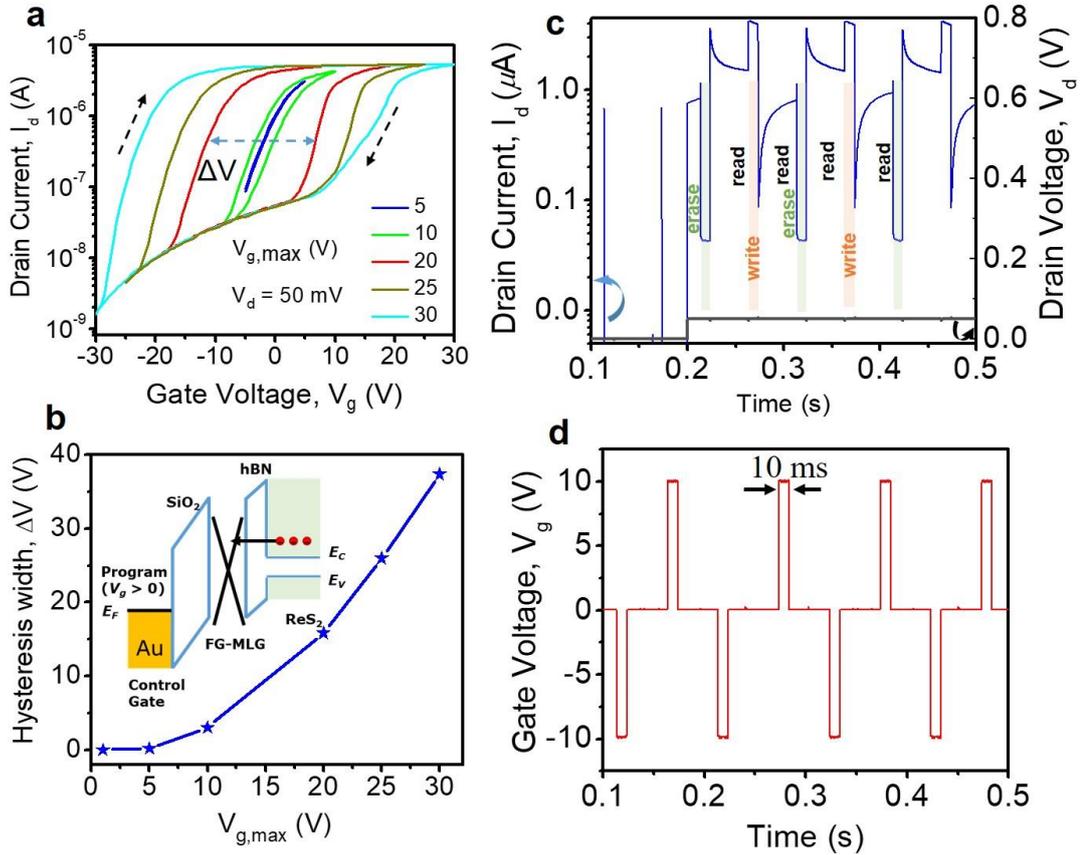

**Figure S12:** Replot of **Figure 4**: Multi-level access via electrostatic coupling: (a) Sweep drain current ($I_d$) versus gate bias ($V_g$) for different gate voltage maxima ($V_{g,max}$) on a logarithmic scale, which exhibits a large memory window. A large memory window enables us to realize multi-level operation. (b) Hysteresis width (i.e. memory window) versus $V_{g,max}$. The inset shows the band alignment across the heterostructure of Au/SiO$_2$/graphene/h-BN/ReS$_2$ under a positive gate bias. Successive program/erase operations using an arbitrary gate pulse of +/- 10 V with a pulse width of 10 ms: (c) $I_d$-t response of the NVM device on a logarithmic scale under a drain bias condition of 0 V for 0.2 s and 0.05 V for the next 0.3s, when using an arbitrary gate pulse of +/- 10 V with a pulse width of 10 ms as shown in (d).

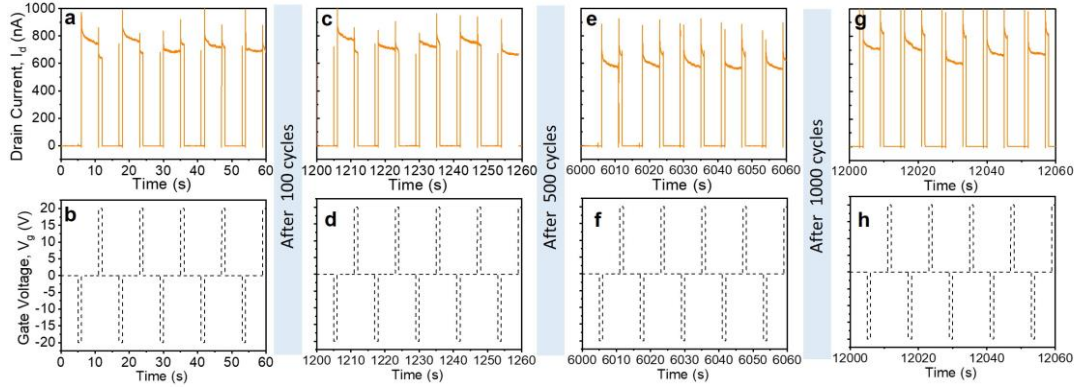

**Figure S13:** Data reliability tests for ReS$_2$/h-BN/graphene memory device using fixed 60 mV drain bias and +/- 20 V gate with 1s pulse width. Cyclic endurance measurements: time-dependent drain current measurements for 5 cycles (a) starting first 5 cycles, (c) after 100 cycles, (e) after 500 cycles and (g) after 1000 cycles under the pulse gates as shown in (b), (d), (f) and (h), respectively.

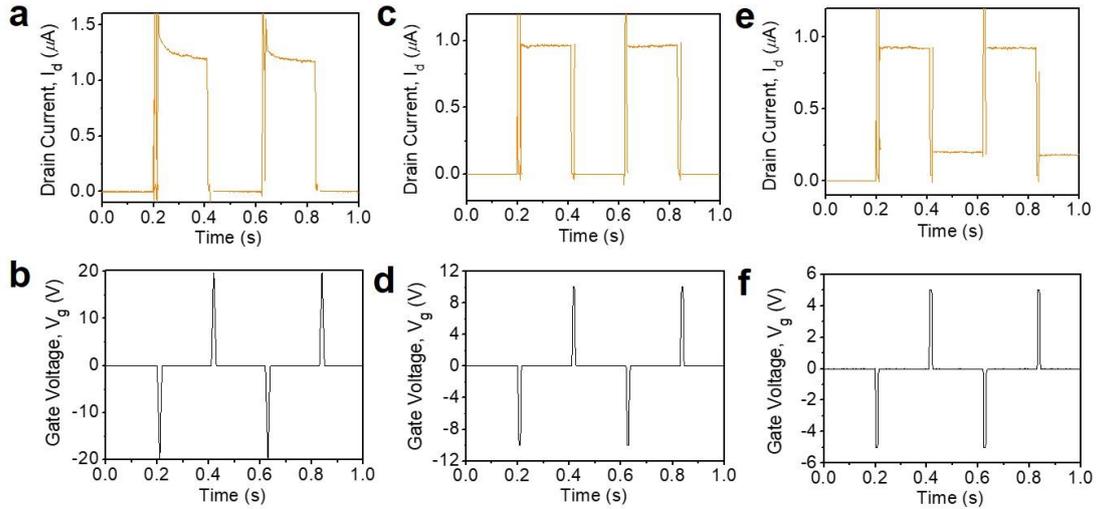

**Figure S14.** Gate voltage dependent NVM operation at fast-time response of 10 ms: $I_d$-t response of the NVM device under write and erase gate pulses of width 10 ms at fixed drain bias of 0.1 V for gate pulse amplitude of (a) 20 V, (c) 10 V and (e) 5 V. Corresponding gate pulses ($V_g$ *versus* time), which were used are plot in (b), (d) and (f), respectively. High and low current level represent write and erase state, respectively.

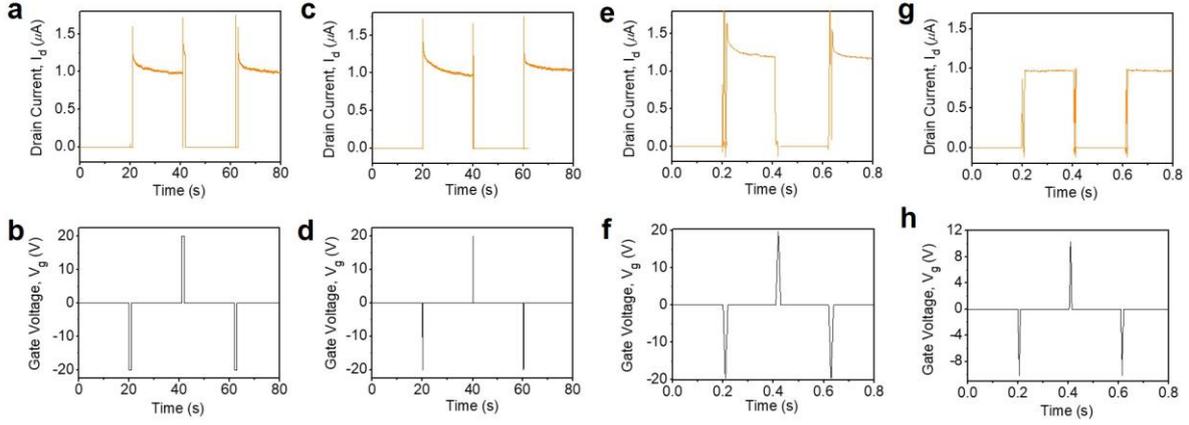

**Figure S15.** Fixed gate pulse width dependent NVM operation at fixed gate bias: $I_d$-t response of the NVM device under write and erase gate pulses of amplitude 20 V at fixed drain bias of 0.1 V for gate pulse width of (a) 1000 ms, (c) 100 ms, (e) 10 ms and (g) 5 ms. Corresponding gate pulses ($V_g$ *versus* time), which were used are plot in (b), (d), (f) and (h), respectively. High and low current level represent write and erase state, respectively.

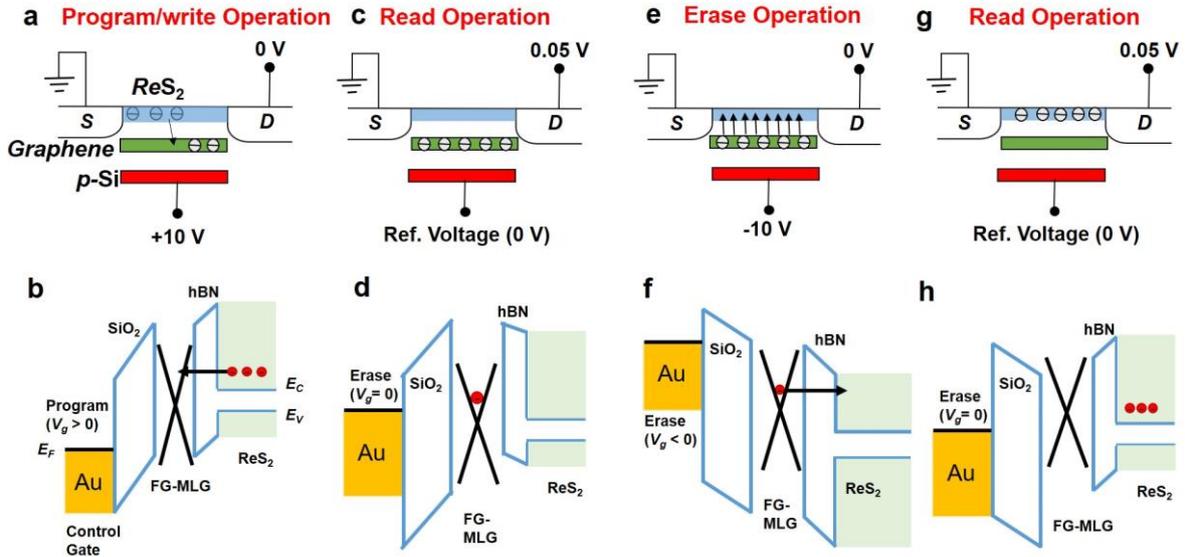

**Figure S16.** (a,b) Program/write operation and band alignment of the $ReS_2$/h-BN/FG-graphene heterostructure device under positive control gate bias and zero drain bias. In this trapping process, free electrons from $ReS_2$ channel move to graphene. (c,d) Read operation and band alignment of the device under the condition of zero gate bias and low drain bias. (e,f) Erase operation and band alignment under negative gate bias and zero drain bias

conditions. In this releasing process, the trapped electrons in graphene move back to the ReS$_2$ channel. (g,h) Read operation after erase process.

| Parameters | Laser-assisted Multilevel 2D Flash Memory |
|---|---|
| Channel material | ReS$_2$ |
| Barrier | h-BN |
| Charge storage/floating gate | Graphene |
| Carrier mobility (cm$^2$/Vs) | 10 – 20 |
| On/off current ratio | 10$^3$ – 10$^4$ |
| Memory window | 38 V from ±30 V |
| Operating power | V$_d$ = 50 mV  <br> Pulse = ± 10 V |
| Endurance | >1000 cycles |
| Retention time | > 10$^4$ s |
| Sample size | Lateral size < 5 μm <br> Thickness ~ 10 nm |
| Laser used for multilevel | Wavelength 532 nm <br> Intensity ~ 1 - 4 mW/cm$^2$ |

**Table** 1: Various parameters of the laser assist multilevel non-volatile memory device.

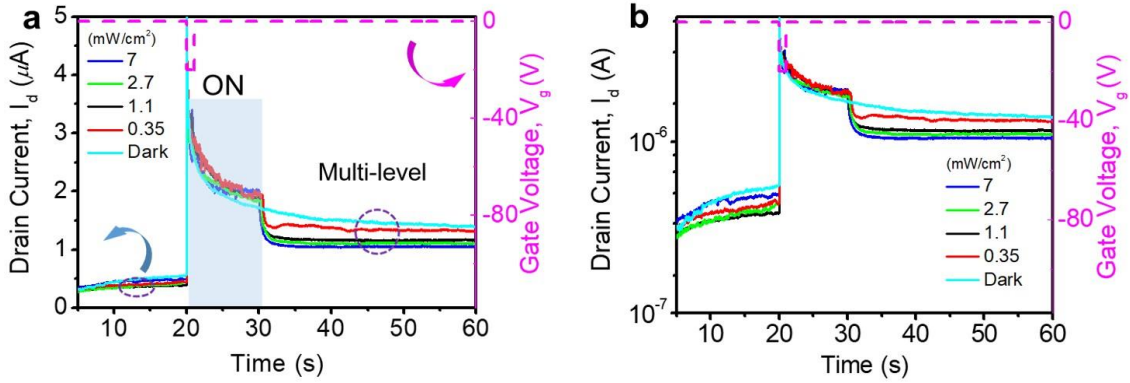

**Figure S17.** (a) Multi-level memory access under varying laser intensity. $I_d$-t response under condition of both laser irradiation as shown by blue shaded area and gate bias as shown by the dotted pink line. Multi-levels corresponding to different ON current states are shown by the circular dotted line. (b) Figure (a) is plotted in logarithmic scale.

**Supplementary Note 2: Different laser intensity tuned multi-level memory operation**

Laser illumination excites the electrons from the valence band into the conduction band of $ReS_2$, and simultaneously the application of negative gate pulses (under erasing conditions of the NVM) helps to recombine the photo-generated electrons. Thus we do not observe a significant difference in photocurrent ($I_P^{ON}$) generation by varying laser intensity under read operation (**Figure S17**). The lower value of read-out current when the laser is OFF ($I_p^{OFF}$) depends on the magnitude of the incident laser intensity. The method to program the NVM using laser pulses for multi-level access as a photoelectric memory device is as follows. The photocurrent, $I_p$, decreases instantaneously after removing the laser pulse and the current reaches a stable storage current ($I_s$). Different light intensities produce different $I_s$, which lead to access different memory states in the device as shown by the dotted circle in **Figure S17**.

**References:**

[s1] J. Y. Park, H. E. Joe, H. S. Yoon, S. H. Yoo, T. Kim, K. Kang, B. K. Min, S. C. Jun, *ACS Appl. Mater. Interfaces* 2017, 9, 26325−26332.
[s2] J. Jiang, C. Ling, T. Xu, W. Wang, X. Niu, A. Zafar, Z. Yan, X. Wang, Y. You, L. Sun, J. Lu, J. Wang, Z. Ni, *Adv. Mater.* **2018**, 30, 1804332.